# Coexistence of Multiple Magnetic Interactions in Oxygen-Deficient $V_2O_5$ Nanoparticles

**Tathagata Sarkar[1], Soumya Biswas[1], Sonali Kakkar[2], A. V. Raghu[1], S. D. Kaushik[3], Chandan Bera[2], Vinayak B. Kamble[1,*]**

[1]School of Physics, Indian Institute of Science Education and Research Thiruvananthapuram, India 695551.

[2]Institute of Nano Science and Technology, Habitat Center, Phase-X, Mohali, Punjab 160062, India.

[3]UGC – DAE Consortium for Scientific Research Mumbai Centre, Bhabha Atomic Research Centre, Mumbai-400085, India

E-mail: kbvinayak@iisertvm.ac.in

**Abstract.** This paper reports on the spin glass-like coexistence of competing magnetic orders in oxygen-deficient $V_2O_5$ nanoparticles having a broad size distribution. X-ray photoelectron spectroscopy yields the surface chemical stoichiometry of nearly $V_2O_{4.65}$ due to significant defect density. Temperature-dependent electrical conductivity and thermopower measurements demonstrate a polaronic conduction mechanism with a hopping energy of about 0.112 eV. The $V_2O_{5\text{-}\delta}$ sample exhibits strong field as well as temperature-dependent magnetic behaviour when measured with a SQUID magnetometer, showing positive magnetic susceptibility across the temperature range of 2-350 K. Field-cooled and zero-field-cooled data indicate hysteresis, suggesting glassy behaviour. The formation of small polarons due to oxygen vacancy defects, compensated by $V^{4+}$ charge defects, results in Magneto-Electronic Phase Separation (MEPS) and various magnetic exchanges, as predicted by first-principle calculations. This is evidenced by the strong hybridisation of V orbitals in the vicinity of vacant oxygen site. An increase in $V^{4+}$ defects shows an antiferromagnetic (AFM) component. The magnetic diversity in undoped $V_2O_{4.9}$ originates from defect density and their random distribution, leading to MEPS. This involves localised spins in polarons and ferromagnetic (FM) clusters on a paramagnetic (PM) background, while $V^{4+}$ dimers induce AFM interactions. Electron paramagnetic resonance spectra measured at different temperatures indicate a dominant paramagnetic signal at a g-value of 1.97 due to oxygen defects, with a broad FM resonance-like hump. Both signals diminish with increasing temperature. Neutron diffraction data rules out long-range magnetic ordering, reflecting the composition as $V_2O_{4.886}$. Despite the FM hysteresis, no long-range order is observed in neutron diffraction data, consistent with the polaron cluster-like FM with MEPS nature. This detailed study shall advance the understanding of the diverse magnetic behaviour observed in undoped non-magnetic systems.

## 1. Introduction

The origin of ferromagnetism in otherwise non-magnetic ($d^0$) materials has been an intriguing question for about a few decades.[1–8] Various reports have shown the existence of unexpected ferromagnetism (FM) or even other less ordered types of magnetism such as anti-ferromagnetism (AFM), paramagnetism (PM) and superparamagnetism (SPM) etc, in simple as well as complex oxides, nitrides which are otherwise expected to be diamagnetic (DM). *Primafacie* the origin of the unusual magnetism was ascribed to magnetic ion dopants in host oxides, but later they were identified as vacancy defects in lattice which contribute reasonably long-range order to give ferromagnetic moment even in undoped samples at room temperature[2, 5, 9–12]. Subsequently, the observations of defect-induced magnetism were reported in other non-oxide systems [13–17] as well as recently in graphene and its analogues[18–21]. However, the net concentration and distribution of these defects (vacancies) is contingent upon the processing condition, hence diverse nature of magnetic properties was reported for the same system in the literature which made it difficult to pin down the origin of this phenomenon.

Efforts were made by various researchers to induce magnetic moments or enhance the existing moment in non-magnetic oxides through doping by transition metal ions. Those have been explained through the Bound Magnetic Polaron (BMP) model where magnetic exchange happens between two polaronic sites through transition metal ions[22–24]. However, this could not be applied to undoped non-magnetic oxides since there were no magnetic impurities to mediate. Nevertheless, there seems to be a larger consensus on the defect at one of the constituent sites being responsible for this unusual magnetism in undoped systems. The transition metal oxides may show various sub-stoichiometric magnetic phases and thus several magnetic properties are depicted due to variable *d*-orbital occupancy. This results in a wide variety of magnetic signatures in oxides of the same element. Nevertheless, the most stable transition metal oxides (TMO), having empty d orbitals ($d^0$, S=0), are expected to show bulk diamagnetism because of their zero net spins e.g. vanadium pentoxide ($V_2O_5$), titanium dioxide ($TiO_2$) etc. Among these, $V_2O_5$ due to its narrow band gap (~2.2 eV) has attracted significant interest for visible absorption as opposed to other optically transparent wide bandgap oxides ZnO, $SnO_2$, $TiO_2$, which have been extensively studied for dilute magnetic semiconductors[6, 9, 12, 22, 25–27]. Moreover, $V_2O_5$ has a layered crystal structure and hence can be exfoliated into atomically thin layers. This has far-reaching implications for 2D materials-based functionalities. Quite a few researchers have reported the observation of magnetic interactions in the vanadium oxide ($V_2O_5$) system. However, like other oxide systems, the magnetic behaviours observed are quite diverse in this case too.[28– 35] Moreover, these studies have reported a wide range of micro-structures of $V_2O_5$, such as nanotubes, nano-lamellas, nanowires, nanoparticles etc.

Mesoscopic materials form an integral part of condensed matter systems and polydispersity is inherent to real materials that cannot be avoided in most generic cases even including the growth of single crystals where a size distribution is inevitable. Measurements performed on such single crystals are representative of the entire batch of samples and caution should be exercised while reporting the phenomena with relevant statistical variations. Nevertheless, such statistical variations are small when reporting macroscopic or bulk properties where the magnitude is large enough. In the case of results about non-trivial origins such as noise, inhomogeneities, crystalline imperfections etc, the statistical variations become overwhelmingly large. Nevertheless, their macroscopic average consistently behaves in a certain fashion whose magnitude may vary from sample to sample and depending on different processing conditions[34] To have a complete understanding of what is the breadth of the different interactions possible, it requires marking the limiting cases at extremes of completely homogenous vis-à-vis an inhomogeneous systems. A system which contains large inhomogeneities is likely to exhibit all possible phenomena which are otherwise reported in diverse studies with a limited degree of inhomogeneity in different compartments. [36–38]

Investigation of the defect-induced magnetism in otherwise non-magnetic oxides has led to new questions and new possible applications. [25, 26, 39] Besides, recently exfoliated two-dimensional layered solids showing different physical attributes are being explored for new physics and devices [40, 41]. However, making magnetic measurements in individual 2D layers is challenging. Therefore, bulk solids like $V_2O_5$ which have layered structures are interesting systems to investigate for this dual significance. This being the case, we have explored the magnetic properties of $V_2O_5$ nanoparticles, which are bulk diamagnetic.

The experimental results were analysed in the light of first principle calculations which support the observed nature of diverse magnetic interactions. Although the BMP model is developed for magnetic impurity doped semiconductors, we find that it also applies to systems which are self-doped with magnetic ions due to charge nonstoichiometry. Further, for the first time, the direct estimation of electric polaron properties has been made in the same system in addition to magnetic measurements. The results of this work demonstrate that there exists a mix of various magnetic interactions in the nanoparticles of $V_2O_5$, which are field as well as temperature-dependent. Thus, it may address the origin of different magnetic phases reported in $V_2O_5$ for different processing conditions and dimensionalities.

The manifestation of inhomogeneous magnetic properties due to short-range phase inhomogeneities is referred to as magnetoelectronic phase separation[42, 43]. Usually, it is observed that upon carrier injection certain systems show FM clusters in non-FM matrix leading to complex magnetic behaviour as demonstrated by several systems. The examples include cobaltites ($La_{1-x}Sr_xCoO_3$[44]), manganites ($Nd_{0.5}Ca_xSr_{0.5-x}MnO_3$[45], $La_{0.7}Ca_{0.23}Sr_{0.07}MnO_3$ [46]) etc. Notably, the transport properties of these systems have a commonality of the polaron hopping mechanism. Similarly, in oxygen-deficient $V_2O_5$ one can imagine electrons doped by oxygen vacancy dopants and their polaron formation. In such cases, the magnetic as well as the transport properties are governed by the site occupancy-induced randomness of the dopant (defect in this case) vis-a-vis the background spin distribution. Thus, we observe a possible spin or cluster glass system wherein competing magnetic interactions are observed having their origin in electronic localisation-inducing magnetoelectronic phase-separated clusters. [36, 47]

## 2. Experimental details

### *2.1. Synthesis and characterisation of nanoparticles*

The precursor, $NH_4VO_3$, in powder form was heated in an alumina crucible at 500$^o$C for about 3 hours (heating rate 10$^o$C/min.) under an ambient atmosphere, in a tube furnace. By 500 $^o$C the thermal decomposition reaction was complete and pure $V_2O_5$ was left behind in the crucible. $NH_3$ as well as $H_2O$ are evolved as gaseous byproducts. The
details of thermochemical reactions can be found elsewhere.[48, 49]

The X-ray diffraction patterns were recorded on a Bruker D8 Advance diffractometer with Cu $K_\alpha$ source having a wavelength of 1.5418 Å. The patterns were recorded in the range 20$^o$ - 80$^o$, with a step size of 0.02$^o$. TEM images were obtained on an FEI transmission electron microscope with an accelerating voltage of 300 keV. The magnetic measurements have been performed using a very sensitive Superconducting Quantum Interference Device (SQUID) magnetometer (Quantum Design, model MPMS3) in the temperature range of 2 to 350 K. The resistivity was measured in a home-built setup using four probe van der Pauw method on a pellet sample of 8 mm diameter and 2 mm thickness.

### *2.2. Neutron Diffraction*

To further comprehend structural characterizations and potential long-range magnetic ordering in the title compound, the sample was subjected to neutron diffraction (ND) at room temperature and low temperature (2K). A powder diffractometer (PD-3) at the National Facility for Neutron Beam Research (NFNBR), Dhruva reactor, BARC Mumbai(India) was utilized. This Multi-PSD (position sensitive detector) based focusing crystal diffractometer was installed by UGC-DAE CSR Mumbai Centre for ND study [50]. The powder sample was filled in a cylindrical sample holder which was then placed at the beamline in a Close Cycle Refrigerator based cryostat (M/s. Cryogenics Industries, UK) for room temperature and low temperature (2 K) ND experiment. The monochromatized neutron beam (wavelength 1.48 Å) was utilized for the neutron diffraction experiment and observed neutron diffraction data was refined by the Rietveld method. FullProf suite software was utilized for the data analysis [51].

### *2.3. Computational details*

First-principle calculations using Density Functional Theory (DFT) were performed with the Projector Augmented-Wave (PAW) method as implemented in the Vienna Ab-initio

Simulation Package (VASP) [52–54]. The Generalized Gradient Approximation (GGA) with the Perdew-Burke-Ernzerhof (PBE) exchange-correlation function was used [55]. The electronic and magnetic properties of $V_2O_5$ were investigated for three cases: pure $V_2O_5$, $V_2O_5$ with 10% oxygen vacancies, and oxygen-deficient $V_2O_5$ with 30% of $V^{5+}$ ions substituted by $V^{4+}$ ions. The DFT+U method was employed to account for the strong on-site Coulombic interactions between the localized 3d electrons of the transition metal V, using Dudarev's approach [56], where only the effective U (U-J) is relevant. Fixed values of 3 eV and 0.9 eV were used for U and J, respectively, for the V atoms. Structure optimization was carried out with a gamma-centered k-point mesh for supercell calculations. A kinetic energy cut-off of 400 eV was used for the plane-wave basis set of 4 × 2 × 6. Lattice constants and atomic coordinates were fully relaxed with a convergence criterion for a total energy of $10^{-3}$ eV. Self-consistent charge density calculations were performed with the same k-mesh. Spin-polarized calculations were conducted with VASP to study the magnetic properties.

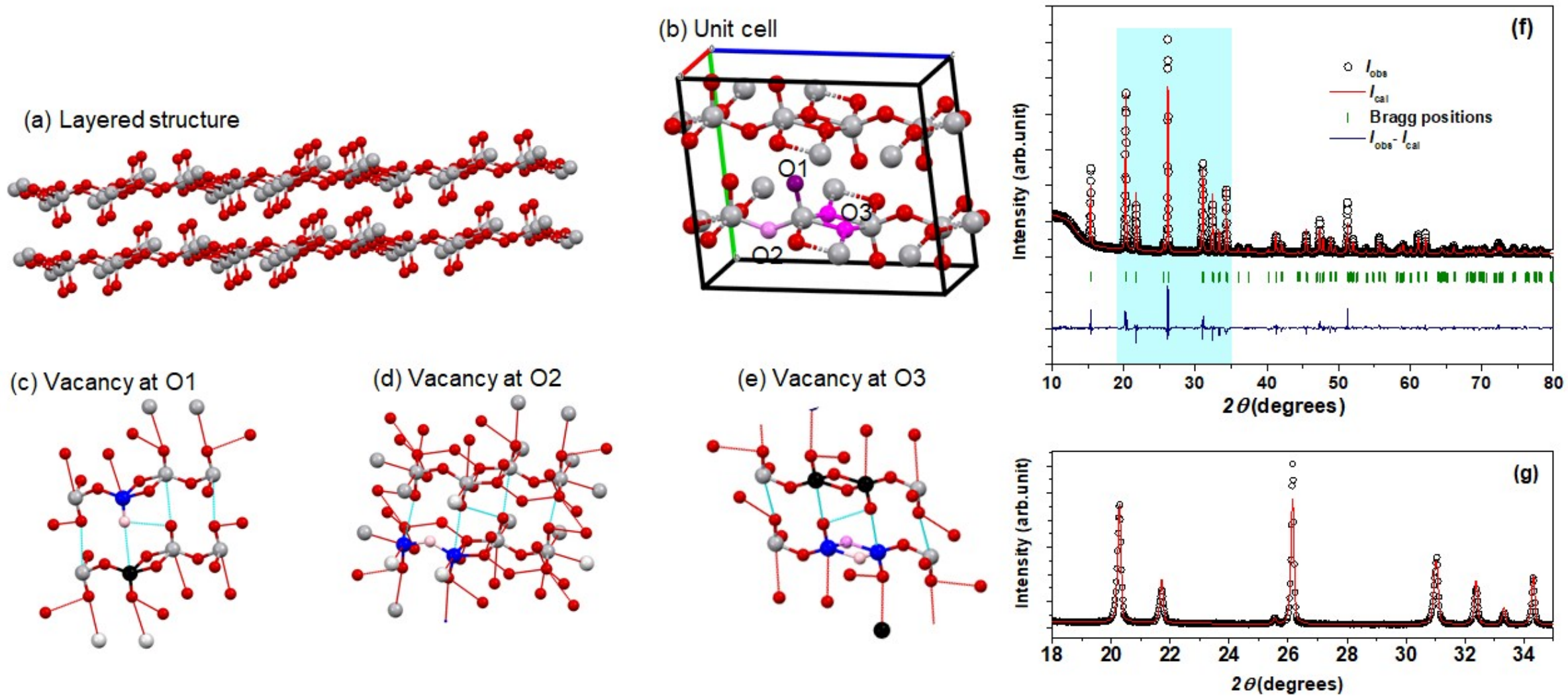

**Figure 1.** (a)The layered crystal structure of $V_2O_5$ has weakly bonded sheets like ladders. (b) The unit cell of $V_2O_5$ and its possible three equivalent sites of oxygen in a $VO_5$ square pyramids. Thus, the three types of oxygen vacancies could arise in the $V_2O_5$ lattice. (c) apical, (d) bridging and (e)doubly degenerate positions (Colour code: Pink: vacancy site, blue: directly bonded to vacancy site, black: weak hydrogen bonded to vacancy site. (f) X-ray diffraction pattern recorded for $V_2O_5$ sample, (g) its magnified view of low angle region showing no impurities.

## 3. Results

### *3.1. Structure, size and crystallinity of $V_2O_5$ nanoparticles*

Vanadium pentoxide has a layered structure wherein parallel sheets of atoms exist along the b-axis as shown in Fig 1(a). It has an orthorhombic structure with space group *Pmmn* and lattice parameters of **a** = 11.51 Å, **b**=3.56 Å, and **c** =4.33 Å as shown in Fig 1(b). As shown here, each $VO_5$ square pyramid is slightly distorted with apical oxygen marked as O1, while the basal oxygen ions can be identified as having one bridging oxygen (O2) and two having equivalent positions (O3). Ngamwongwan et.al.[57] studied the energetic formation of vacancy defects at each of these sites and the corresponding effect on the band structure. It was demonstrated that the formation of a neutral vacancy at any site acts as a double ionised donor with these electrons localisation of vanadium ions. Also, it was shown that out of these possible sites, vacancy at O1 is a major defect for $V_2O_5$ synthesised at atmospheric conditions. Besides, the higher the temperature of annealing or formation, the higher the oxygen defect concentration and subsequently the carrier concentration of the solid. Nevertheless, having an oxygen vacancy at any of the sites causes further lattice distortion in $VO_5$ square pyramids. Besides, the weak van der Waals bonds between the two layers (shown by broken bonds in Fig 1(c-e)) also get affected in adjacent layers due to vacancies.

The XRD pattern collected for $V_2O_5$ is shown in Fig 1(f). The obtained data has been analysed using the Rietveld refinement profile fitting method. The full range of the XRD pattern shows all reflections of the orthorhombic structure (JCPDS no. 600767). The Rietveld refinement parameters including lattice parameters, occupancy etc have been summarised in Table S1 in the Supplementary Materials section. Fig 1(g) shows the magnified view of the low angle region of the XRD pattern, which indicates the absence of impurity phases in the sample, seen to the best of resolution. Further, the crystallite size calculated using Scherrer formula is estimated to be about 50 nm. The size and crystallinity of the particles have been further verified with Transmission Electron Microscopy (TEM). The images obtained have been shown in Fig 2(a). It may be observed from the TEM image that the typical size of the particles seen is nearly 50 to 100 nm, as shown in Fig 2(b). As the XRD shows the lower threshold of the crystallites, these observations are in agreement with each other. Moreover, there is a greater degree of agglomeration within the particles as the thermal decomposition process is more akin to producing agglomerates as opposed to solution-based methods where nuclei grow into well-separated particles[58]. The lattice fringes obtained in high-resolution TEM have been analysed using Fast Fourier Transform (FFT) and the spacing observed corresponds to (010) planes of $V_2O_5$, as marked in Fig. 2(c) and (d). Fringe separation has been measured using FFT of the fringes as shown in Fig. 2(d). The spacing of 0.36 nm is observed in $V_2O_5$. The selected area electron diffraction (SAED) pattern (shown in Fig 2(e)) was obtained and indexed corresponding to the $V_2O_5$ planes, such as (200), (110), (011), (400), (600), (240) and (170). The elemental scan using energy dispersive spectroscopy (EDS) is shown in Fig 2(f) and it shows peaks of only V and O which marks the purity of the sample. Therefore, no other impurity phases or elements have been observed in the sample.

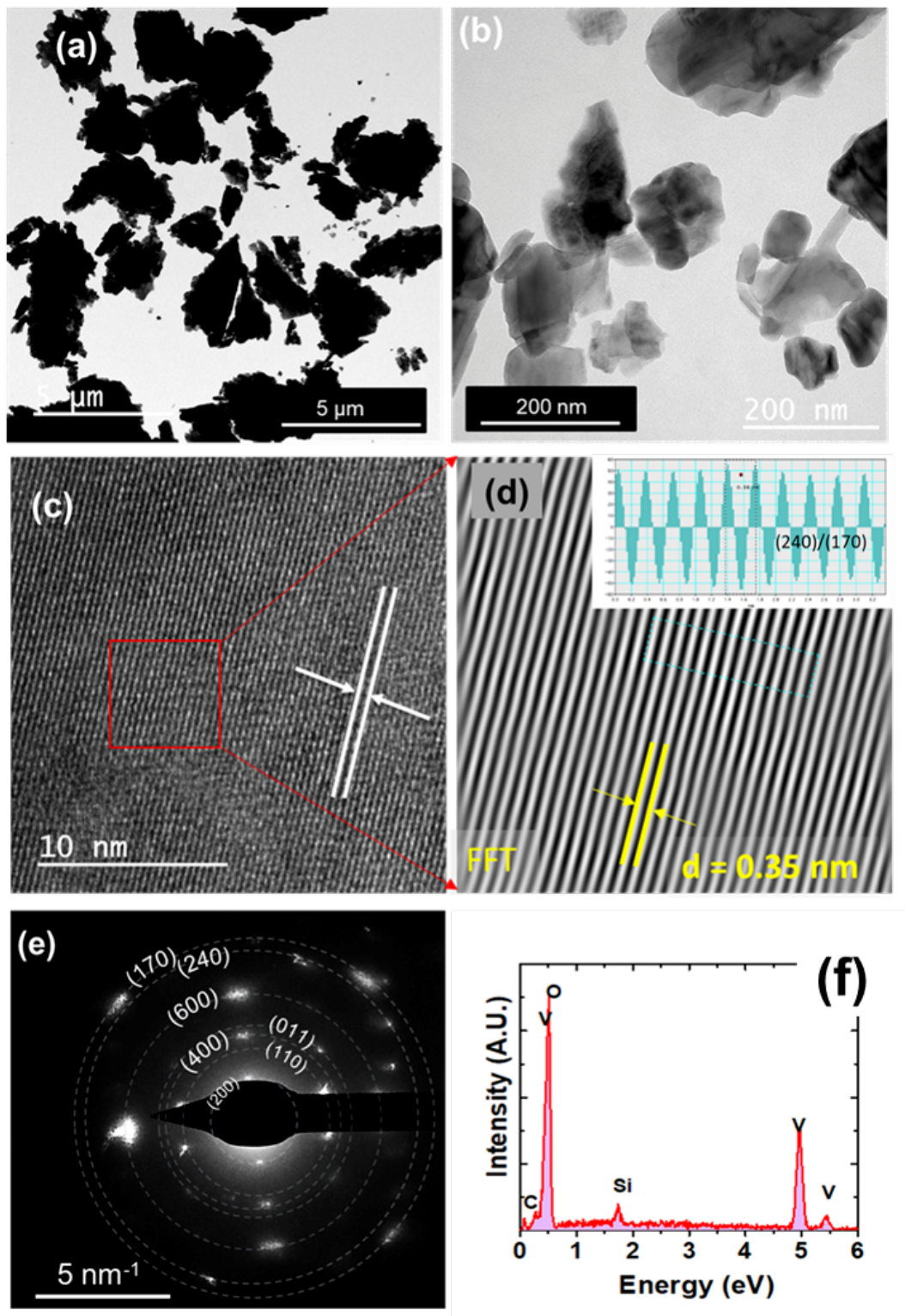

**Figure 2.** Transmission Electron Microscope (TEM) images of the particles at (a) low and (b) high magnifications. (c) High-resolution TEM (HRTEM) image with (d) its Fast Fourier Transform (FFT) for calculating the lattice spacing of 0.35 nm corresponding to the (101) spacing of $V_2O_5$. (e) The SAED pattern collected on the sample (f) Energy Dispersive X-ray Spectrum (EDS) showing only vanadium and oxygen (Si signal from substrate).

The X-ray and UV photoelectron studies of $V_2O_5$ were performed and the data reported in our earlier work[49] is presented for a ready reference. The O 1s peak shows asymmetry with a higher binding energy shoulder ascribed to O ions in the vicinity of oxygen defects in the crystal[9]. The most probable defects are the oxygen vacancies in the $V_2O_5$ lattice, which act as donors and impart an n-type conductivity to wide bandgap oxides[13, 59]. Since XPS is a surface-sensitive technique, these oxygen vacancies are mainly on the surface due to abrupt surface discontinuity in the periodicity of the lattice. Nevertheless, their relative fraction concerning regular O 1s is significant. The XPS is a local as well as superficial probe (with a sampling depth of <10 nm. The same is about 1 $\mu$m for EDS, however it is far less accurate). XPS gives a (local) accurate surface quantification of the composition. It may be only used as a reference. Nevertheless, the sample surface was cleaned by ion milling before collecting the spectrum. These oxygen vacancies are warranted by the thermodynamic stability of the lattice and their formation energy is much lower [57], which justifies their dominant nature as compared to $V^{5+}$ metal ion vacancies. The UV photoelectron spectroscopy (UPS) also shows the presence of a small mid-gap peak, about 1 eV below the VBM, ascribable to the presence of reduced oxidation states of V i.e. $V^{4+}$[13]. The V spectrum shows strong non-stoichiometry of V ions in 4+ and 5+ states. The resolved peaks have been ascribed to chemical shifts due to the occurrence of $V^{4+}$ substitution at $V^{5+}$ sites. The presence of $V^{4+}$ is warranted for charge neutrality, balancing the oxygen vacancies, i. e. lack of negatively charged ions.

**Table 1.** Quantification results from the X-ray photoelectron spectroscopy data.

| | $V_2O_5$ | |
|---|---|---|
| Element | V | O |
| Estimated fraction | 2 | 4.63 |
| Estimated formula | $V_2O_{4.63}$ | |
| $V^{4+}$:$V^{5+}$ | 39.9: 60.1 | |
| $O^{2-}$:$V_O$ | 60.5: 39.5 | |

### *3.2. What do transport measurements reveal?*

The electrical resistivity $(\rho)$ vs temperature $(T)$ data for the $V_2O_5$ sample is obtained using a 4-probe van der Pauw geometry, in the temperature range 160 to 600 K. The data reveals a semiconducting nature of the sample as the resistivity drops with increasing temperature, as shown in Fig 3(a). However, when the Arrhenius plot is obtained i.e. ln $\rho$ vs $1/T$, the activation energy of the sample turns out to be very low 0.11 eV. Such a low value of thermal activation energy for a large bandgap material (~ 2 eV, See supplementary material Fig. S1)) cannot justify the thermally activated behaviour in band transport. In Fig 3(a and b) the sudden break/discontinuity in the resistivity data at 300 K, as shown in the plot. This is just a measurement artefact which resulted from the change of the experimental set-up used while going from the low-temperature data (T < 300 K) to the high-temperature data (T > 300 K). Two different setups were used for the resistivity measurements on the same sample pellet because there is no single setup for measuring resistivity across this entire temperature range. A custom-made close-cycle cryostat based low temperature set-up was used below 300 K while another custom made-heater based high temperature set-up was used from 300 K to 600 K.

This apparent discrepancy and the exact nature of transport of $V_2O_5$ has been ascribed to the low-lying impurity levels and is well documented in the literature [60, 61]. The charge imbalance resulting from oxygen vacancies in $V_2O_5$ is compensated by the excess electrons located on neighbouring vanadium atoms, reducing them from $V^{5+}(d^0)$ to $V^{4+}$ $(d^1)$. This is directly substantiated by the XPS data which indicates the significant presence of $V^{4+}$ in the given sample, as reported in our earlier work [49]. Nevertheless, XPS being a surface-sensitive technique, those are most likely the surface defects formed as a result of surface oxygen vacancies. However, those are also likely to be present in the bulk of the nanoparticles, although in reduced concentrations balancing the net defect density. The localised charge on $V^{4+}$ deforms the lattice in its vicinity and leads to the formation of a bound state of lattice distortion and a trapped electron, namely a polaron. These polarons can carry both current as well as heat as they move within the solid. The transport properties of the $V_2O_5$ nanoparticles are probed as a function of P doping and the same is reported in our earlier work [49].

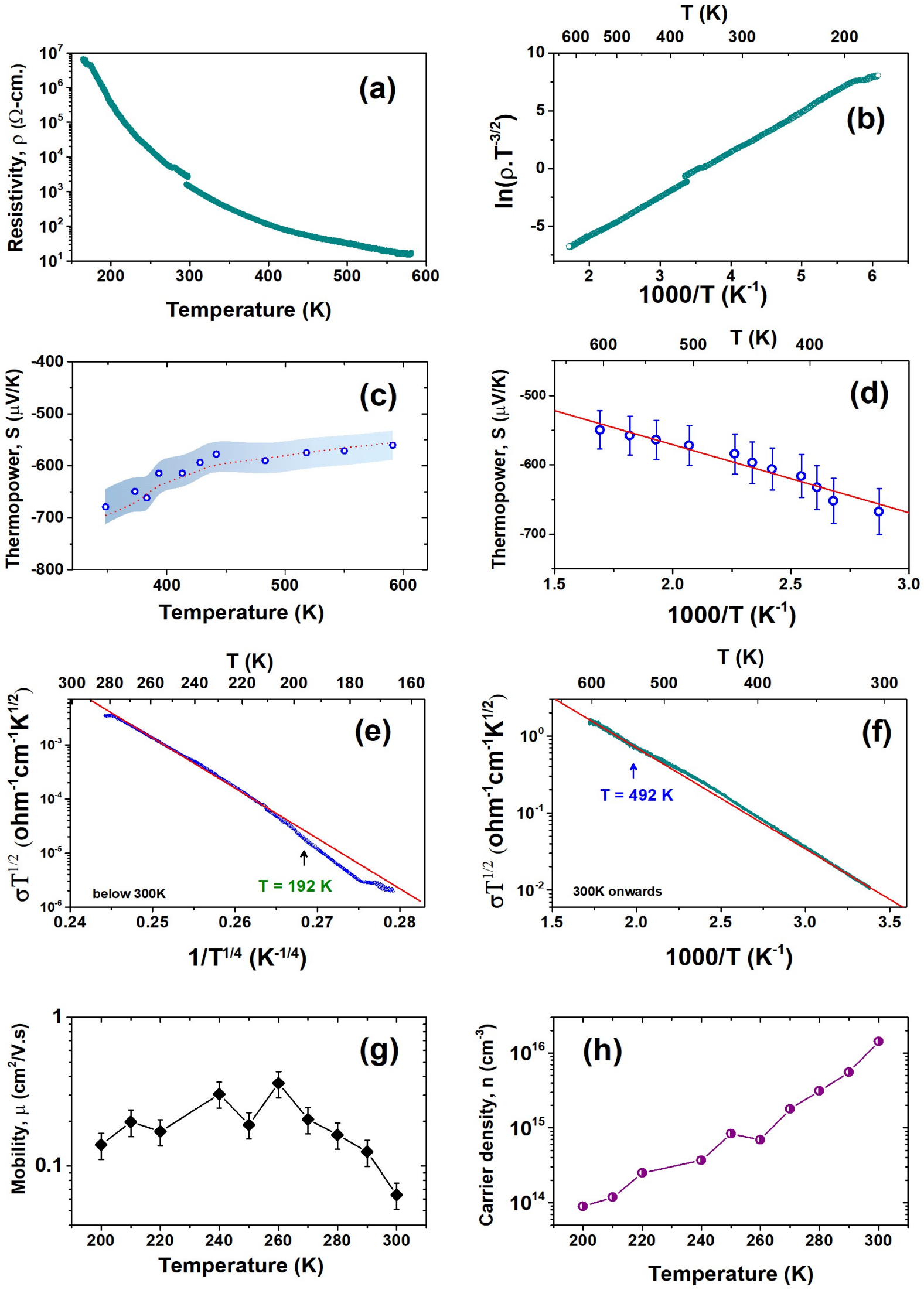

**Figure 3.** (a) Resistivity vs temperature plot of $V_2O_5$ pellet and the fit using (b) polaronic model. (c) Seebeck coefficient vs temperature plot of $V_2O_5$ pellet (the dashed line is a guide for the eye) and (d) the fit using the polaronic model. (e) ln $\sigma T^{1/2}$ vs $1/T^{1/4}$ plot from room temperature to high temperature and (f) ln $\sigma T^{1/2}$ vs $1/T$ from room temperature to low temperature. Measured values of (g) carrier mobility and (h) carrier concentrations using hall measurements.

**Table 2.** The estimated activation energies and their constituents.

| Data | slope value | energy (eV) |
|---|---|---|
| Resistivity | 3581.43 | $\Delta E = 0.308$ |
| Seebeck | 0.098 | $\Delta E_g = 0.196$ |
| $\Delta E_p = \Delta E - \Delta E_g$ | | 0.112 |

The temperature dependence of DC resistivity through the polaron hopping channel is given by,

$$\rho = A\, T^{3/2} \exp\left(\frac{\Delta E}{k_B T}\right) \quad (1)$$

where $\Delta E$ is the activation energy of the transport. Fig 3(b) shows that $\ln(\rho T^{-3/2})$ vs $1/T$ is in very good agreement with the above equation, confirming the polaronic transport and yielding $\Delta E = 0.308$ eV. Further, $\Delta E$ is constituted of $\Delta E_p$ and $\Delta E_g$. Where $\Delta E_g$ is the energy required to put the charge into the polaronic site, while $\Delta E_p$ is the energy required to extract the charge from one polaronic site and propagate it to the next. Since all the polarons are assumed to have the same energies, $\Delta E_g$ can be independently ascertained from the Seebeck coefficient ($S$) vs temperature data by fitting it with the equation,

$$S = \frac{k_B}{e} \exp\left(\frac{\Delta E_g}{2 k_B T}\right) \quad (2)$$

Thus, the variation of the Seebeck coefficient with temperature can be used to extract the value of $E_g$. The Seebeck coefficient (S) vs temperature (T) measurement is shown in Fig. 3(c). The $S$ vs $1/T$ data shown in Fig 3(d) fits very well with this expression, confirming the presence of thermal activation of charges into the polaronic sites and yielding $\Delta E_g = 0.196 eV$. These values are summarised in Table 2. Thus, the polaronic hopping energy calculated in this case is $\Delta E_p = 0.112$ eV which is in good agreement with literature for vanadium pentoxide[57, 60] and other complex oxides which also have $\Delta E_p = 0.1$ eV.

The Heike's formula for $S$ is given by [62, 63]

$$S(T \to \infty) = \frac{k_B}{e} \ln\left(\frac{c}{1-c}\right) \quad (3)$$

where $c$ is the ratio of the number of reduced transition metal ions ($V^{4+}$) to the total number of standard oxidation state metal ions ($V^{5+}$)in the transition metal oxide. The negative sign of the Seebeck coefficient implies n-type carriers.

The Seebeck values obtained from experiments are unusually high considering $V^{4+}$ ion constituents estimated from XPS as well as Eq. 3. The estimation of $V^{4+}$ ions gives c at about 0.01 from S vs T. (see Fig. S2 in supporting information). Here we do understand that XPS being a surface-sensitive technique, gives a surface estimation which could be different from the bulk. In nanoparticles, most defects reside on the surface due to uncompensated bonds and surface states. This implies that despite a large number of defects, only a small fraction participates in conduction.

Now, the polaron hopping conduction, as described by Greaves[62], actually consists of two contributors - the optical modes ($\sigma_{opt}$) and the acoustic modes ($\sigma_{ac}$), so that

$$\sigma = \sigma_{opt} + \sigma_{ac} \quad (4)$$

At higher temperatures, only the optical modes contribute to the conductivity while it is only the acoustic modes that are responsible for conductivity at lower temperatures. However, in the intermediate temperature range, contributions from both modes exist. Fig.3(e) shows the plot of log $\sigma T^{1/2}$ vs $1/T$ in the high-temperature range. In Fig. 3(f), log $\sigma T^{1/2}$ has been plotted against $1/T^{1/n+1}$; where $n$ represents dimensionality. Here, n=3 fitting is shown for intermediate temperatures, however, in the low-temperature range, n=1 gives the best fit implying Effros–Shklovskii variable range hopping [49]. In fact, in the very low-temperature data both the plots show good linear fit. A closer inspection of the high-temperature data (Fig. 3(e)) shows that the data deviates from linearity at around 490 K. Above this temperature, acoustic modes are no longer significant. A similar deviation from linearity is observed in the low-temperature data (Fig. 3(f)), at around 190 K. Below this temperature all the optical modes are quenched. In the intermediate temperature range (190 K $< T <$ 490 K), both modes are present and contribute to the conductivity. All these observations comply with the model proposed by Greaves in [62] and the key temperature values obtained are in good agreement with these kinds of glasses.

Thus, in the system of $V_2O_5$ having significant defect density the transport occurs through equivalent site hopping within kT of $E_F$ at low temperatures, assisted by acoustic phonon (representing bound polarons). This is characterised by a large hopping energy. While at higher temperatures the slope of $E_F$ becomes flat across several kT, leading to band-like conduction between one site to the other having equivalent energies due to significant lattice distortion (representing free polarons). This also leads to saturation of thermopower due to no net change in carrier potential energies during hoping at equivalent sites. The same may be observed in the carrier mobility and carrier concentration data shown in Fig. 3(g) and (h) respectively. The sample shows very low carrier mobility and lower carrier concentrations below room temperature (300K). The carrier concentration quickly escalates through thermal activation, while the mobility gets adversely affected due to a sudden rise in free carriers.

### *3.3. Do undoped $V_2O_5$ nanoparticles exhibit magnetic order?*

The magnetic measurements of the $V_2O_5$ sample are carried out on a sensitive SQUID magnetometer. The background data of the magnetometer without a sample is shown in Fig S3 in the supplementary materials section. Fig. 4 shows all the temperature-dependent magnetic measurement data obtained for $V_2O_5$. The M-T measurements are done over the temperature range 2 - 350 K, in three different bias fields, 20, 100 and 1000 Oe. The Zero Field Cooled (ZFC) curves were obtained by cooling the sample in the absence of a magnetic field, up to 2 K. Subsequently, a small bias field was applied and measured in the warming cycle. Field-cooled (FC) curves were obtained by taking measurements while cooling the sample down to 2 K, under a constant bias field. The sample holder the background has been subtracted from all the sample data.

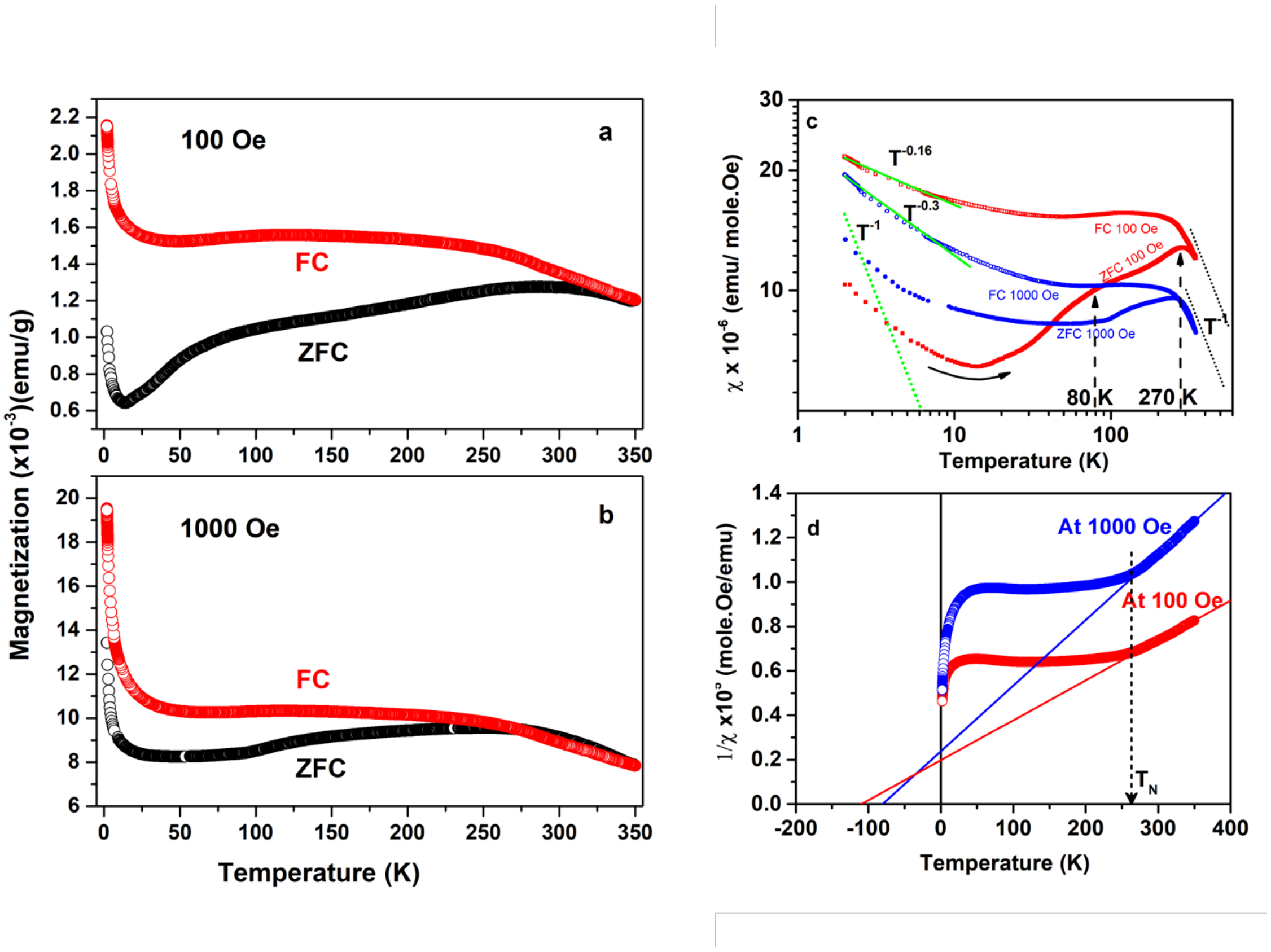

**Figure 4.** (Colour online) (a), (b) M-T (FC-ZFC) at bias fields 100 and 1000 Oe respectively. (c) The DC magnetic susceptibility ($\chi$) vs temperature at 100 and 1000 Oe of undoped $V_2O_5$ nanoparticles. The regions having paramagnetic dominance can be corroborated by comparing them with the $T^{-1}$ line shown for ready reference. The peak in ZFC near 200 K is marked by an arrow. (d) The Curie-Weiss plots for the two field values. The negative intercepts denote antiferromagnetic ordering in the undoped $V_2O_5$ nanoparticles.

The M-T data resembles cases where there is a coexistence of paramagnetic (PM), and ferromagnetic (FM) components within the same material. The FC data at 100 and 1000 Oe shown in Fig. 4 (a) and (b) respectively, have a similar nature, consisting of a rising trend when cooled from room temperature (300 K), a plateau in the intermediate temperature range and again a sharp rise below 10 K. The FC data shows a rise in the moment in a Curie-Weiss (CW) fashion followed by a broad maximum as well as an upturn. These observations are similar to those observed due to the presence of short-range interactions in a glassy system[64, 65]. Towards the lowest end on the temperature scale, a PM-like exponential tail is seen which signifies the PM component of the system. At low temperatures, the sudden fall in moment with temperature is reminiscent of perfect PM behaviour which may be attributed to superparamagnetism (SPM) given the size of the $V_2O_5$ nanoparticles (<100 nm). Similar variations of PM with FM and glass-like FC-ZFC are found in a number of non-magnetic oxide nanoparticles [9, 22, 46, 66–68] The irreversibility of the ZFC and FC curves reveals a spin-glass nature of the system. The gap begins to close at higher fields. That might be due to the dominant PM contribution at higher fields which rises linearly with the field as opposed to FM that saturates. Such behaviour has been previously reported for manganites by Pillai et al. [45] which also associates such behaviour with the inhomogeneous mixing of FM-AFM phases. The ZFC-FC data measured at 20 Oe bias field is shown in Fig. 5 wherein the curve is strikingly different in comparison to those of high field data. Nevertheless, when observed in 20 Oe data it may be noted that as the field decreases (at low temperatures), M-T gradually changes slope with the field. At 1000 Oe it is exponentially decreasing with a steep negative slope, while the slope becomes less negative for 100 Oe and it becomes positive for 20 Oe. This may be interpreted as the suppression of the PM contribution as the field decreases below 100 Oe. The presence of the AFM component is further accentuated in the slowly falling nature of the ZFC curve, where two AFM components are seen, one at 270 K and the other at 80 K[45]. See Fig. S4 in the supplementary material section for a comparison of all the M-T data.

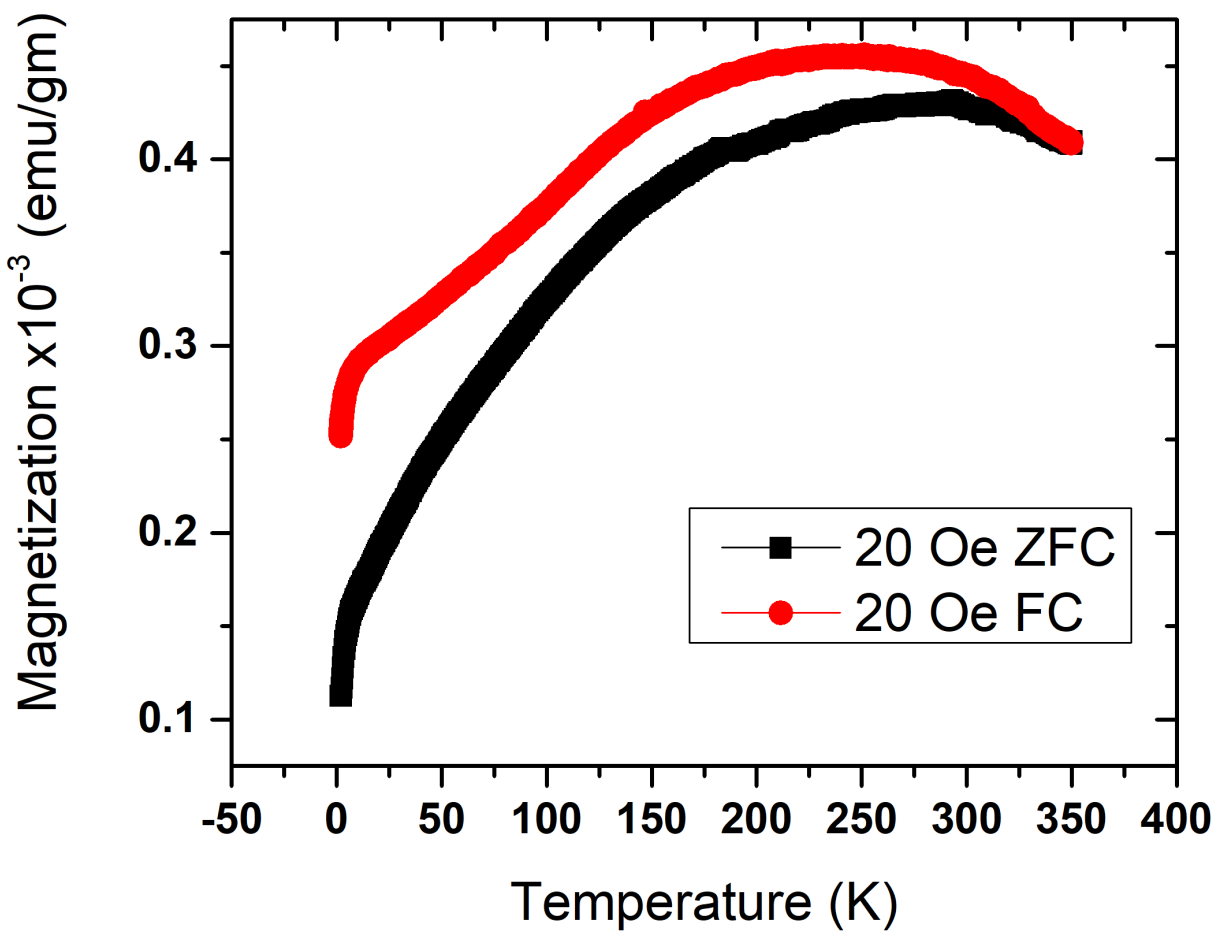

**Figure 5.** M-T (FC-ZFC) at bias fields 20 Oe bias field for undoped $V_2O_5$ nanoparticles.

The DC magnetic susceptibilities are evaluated at given magnetic fields, and the same are plotted in Fig. 4(c). The gradual change in trend with temperature, as mentioned above, is observed in the case of magnetic susceptibilities ($\chi$), shown in Fig 4(c). The temperature dependence systematically reduces the exponent as H increases from 100 to 1000 Oe as shown with green dashed lines. The departure between FC-ZFC is maximum for 100 Oe field, which is intermediate of low (20 Oe shown in Fig. 5) and high (1000 Oe) fields, demarcated in this study, where the susceptibility of the FM centres is gradually surpassed by that of the PM centres and going through a maximum glassy nature in between.

$$\chi = \frac{C}{T-\theta_{CW}} \tag{5}$$

A Curie-Weiss (CW) fit (Eq. 5) is attempted on the high-temperature region of the $1/\chi$ vs T plot, as can be seen in Fig. 4(d), to ascertain the nature of the various magnetic phases predicted in the system. However, the CW extrapolation lines in the high-temperature region, yield negative intercepts on the temperature axis indicating a possible antiferromagnetic (AFM) ordering. Further, the value of CW temperature ($\theta_{CW}$) decreases as the applied field increases. This could be extrapolated as $\theta_{CW} \rightarrow 0$ as H increases (purely PM at high fields). Nevertheless, the apparent Neel temperature is nearly the same irrespective of field i.e. 270 K. Similar AFM ordering has been observed by Dhoundiyal et.al. [35] and Dreifus et.al. [69]. However, the Neel temperatures reported are quite different (50 K [35] and 80 K [69] respectively). These AFM interactions may result from $V^{4+}$ ions, which has been reported in the literature[70, 71]. However, as discussed earlier, the concentration and relative proximity of $V^{4+}$ ions depend on that of the oxygen vacancies which are predominantly governed by the processing conditions as well as microstructures, which are different in these cases. A strong deviation of CW fit for low field values indicates the presence of short-range clusters of FM in AFM or vice versa in glassy systems [64, 65]. Thus, the deviation from CW fit observed for low field (20 and 100 Oe) points towards such short-range FM clusters in the AFM matrix and resulting complexities.

The effective moment is deduced from the Curie constant using the Eq 6,

$$\mu_{eff} = \sqrt{\frac{3k_B C}{N\mu_B}} \tag{6}$$

where N is Avagadro's number, $\mu_B$ is Bohr magneton and C is the curie constant deduced from the CW plot whose value has been found to change with the field. However, as can be seen in Fig 4(b), the 1000 Oe field data shows clear signature of linear behaviour at high temperature, considering that Curie constant value of 0.31 K.emu/mole(of V).Oe which gives an effective moment of 1.56 $\mu_B$ which is close to 1.72$\mu_B$, a value obtained by a single unpaired electron (S=1/2). The comparison of Langevin and Brillouin function fitting to the 2 K M-H data is shown in Fig. S6 and S7 in the supplementary material.

The observations from the M-T data are corroborated by the M-H isotherms, obtained at three different temperatures, 2, 100, and 300 K, as shown in Fig 6(a), (c) and (e) respectively, and their corresponding magnified images in Fig 6(b), (d) and (f) respectively. The high-temperature isotherms are more or less like FM, with a prominent hysteresis loop and saturation, overlaid with a small positive slope, which might come from both PM and AFM components in the system. However, at 2 K there could be likely an SPM component that completely overshadows the FM loop, as expected from the M-T data. Nevertheless, FM has been found to persist from 2 K to 300 K, as depicted from the centred hysteresis loops in Fig 6. Thus, a mixed contribution from various FM-AFM-PM states is observed each of them evolves independently with temperature. The respective contributions have been attributed to various origins discussed in subsequent sections.

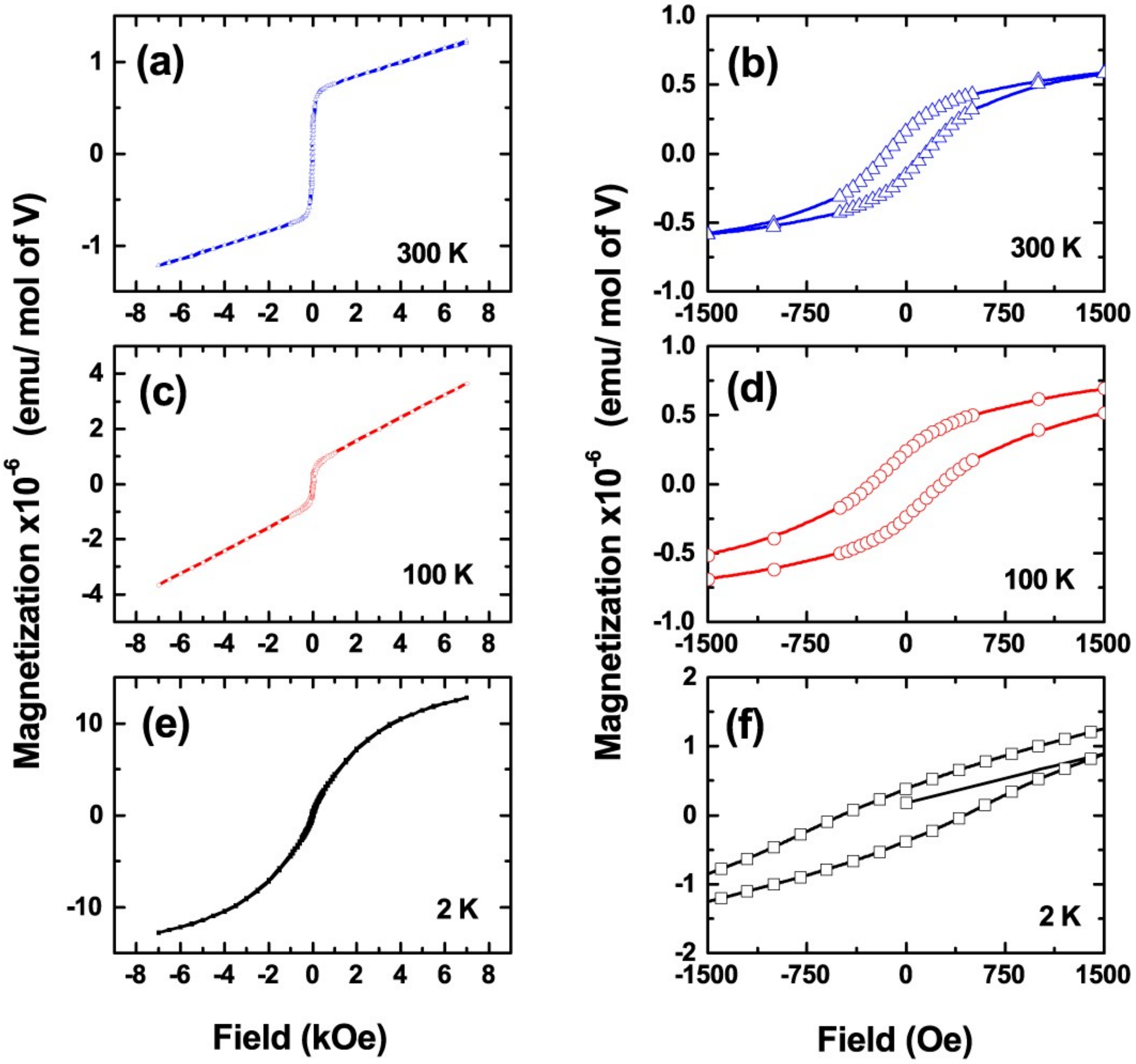

**Figure 6.** Colour online) (a), (b), (c) M-H isotherms $V_2O_5$ at 300, 100 and 2 K respectively. (d), (e), and (f) are the magnified M-H isotherms to show the presence of hysteresis at all temperatures.

Here FM dominates at low fields (depicted by hysteresis loops) and high temperatures, while PM dominates at low temperatures (< 10 K) and high fields (as seen by linear rise of M-H). At the intersection i.e. low temperature, low fields the system behaves like superparamagnetic single-domain particles. However, a system at a low-temperature PM phase turning into an FM at a high temperature is unlikely hence all the components must have a coexistence where the measurements manifest their behaviour depending on the relative strength of magnetic susceptibilities. There could be a wider PM background with short-range clusters of FM and AFM which reveal depending on temperature and field and also convert to PM when heated beyond 300 K.

There is a hysteresis but no saturation is observed till the highest field is applied (7 Tesla). The presence of hysteresis in an M-H isothermal curve without saturation may come from multiple factors like two or more competing phases. The M-T data in the low-temperature range also shows a steep $T^{-0.16}$ and $T^{-0.3}$ dependence for 100 Oe and 1000 Oe data respectively (See Fig. 4(b)) while M-H depicts a hysteresis at all temperatures. Here, because of the wider size distribution of the nanoparticles some of them which are near the lower bound of the distribution may exhibit SPM. (The experimental signatures of likely SPM are compared with the reports in the literature [72] and presented in the supplementary information section E, Fig. S5) Such smaller particles may have much lower curie temperature than the larger size counterparts. The dependency of Curie temperature on size is given by eq [73],

$$T_c(D) = T_c(\infty)\left(1 - \frac{3L}{2D}\right) \quad (7)$$

where D is the size of the nanoparticles, L is the thickness of the monolayer, and $T_C(D)$ and $T_C(\text{inf})$ stand for the Curie temperature of nanoscale and bulk systems respectively. Therefore, it may be possible that the smaller particles may be exhibiting that phase transition to PM much before the larger counterparts.

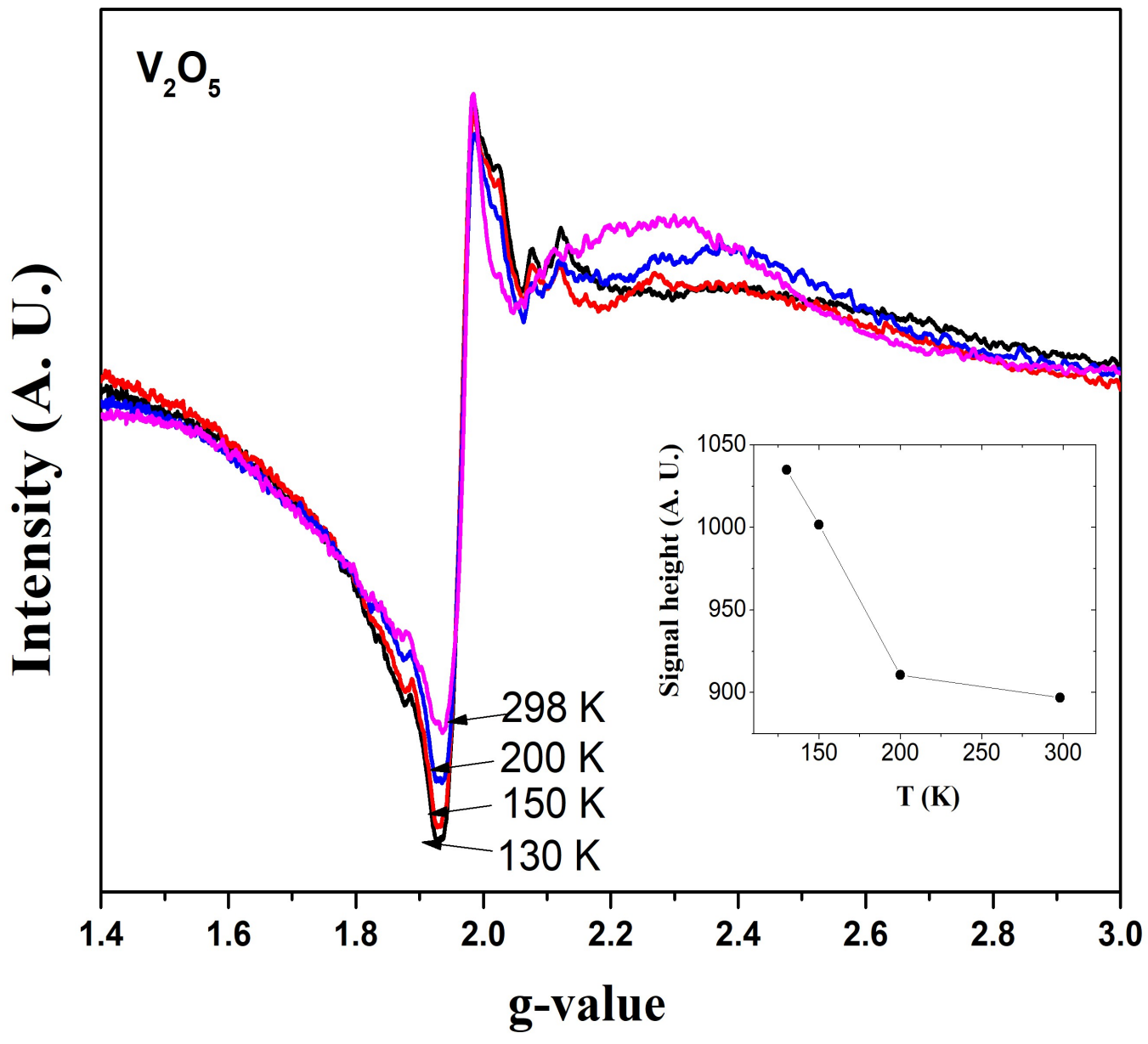

**Figure 7.** EPR spectra of $V_2O_5$ nanoparticles were obtained at four different temperatures viz. 130, 150, 200 and 298 K. The inset shows the signal height vs temperature of the 1.97 g-value signal.

To confirm the nature of magnetic interactions, Electron paramagnetic resonance (EPR) spectroscopy was performed till 130 K the lowest possible temperature attainable in the given system. We have collected EPR spectra at four different temperatures, 130, 150, 200 and 300 K as shown in Fig 7. The spectrum consists of a sharp single line (S=1/2) at a g-value of 1.97 with details of hyper-fine features (corresponding to spin 7/2) barely visible appearing superimposed on the single sharp line. As reported for EPR spectra of single crystals of $V_2O_5$ samples, the 1.97 signal is attributed to oxygen vacancy-induced localised spins. [74, 75]. The signal gets sharper when cooled as shown in the inset figure of signal height vs temperature, depicting a stronger paramagnetic localisation at low temperatures. This is in good agreement with that of the magnetic measurement studies, that show the dominant paramagnetic slopes at low temperatures in M-H data. In our earlier work [49], we have also shown that the feature at a g value of about 2.4 the hump observed may be attributed to the ferromagnetic exchange present in the system. The broad FMR signal appears to be superimposed with the sharp PM signal. The earlier work demonstrated changes in the dominant contribution of PM with increasing P doping in EPR as well as magnetic measurements due to localisation effects as a result of P doping. The signal shows a systematic change in temperature that may be further investigated.

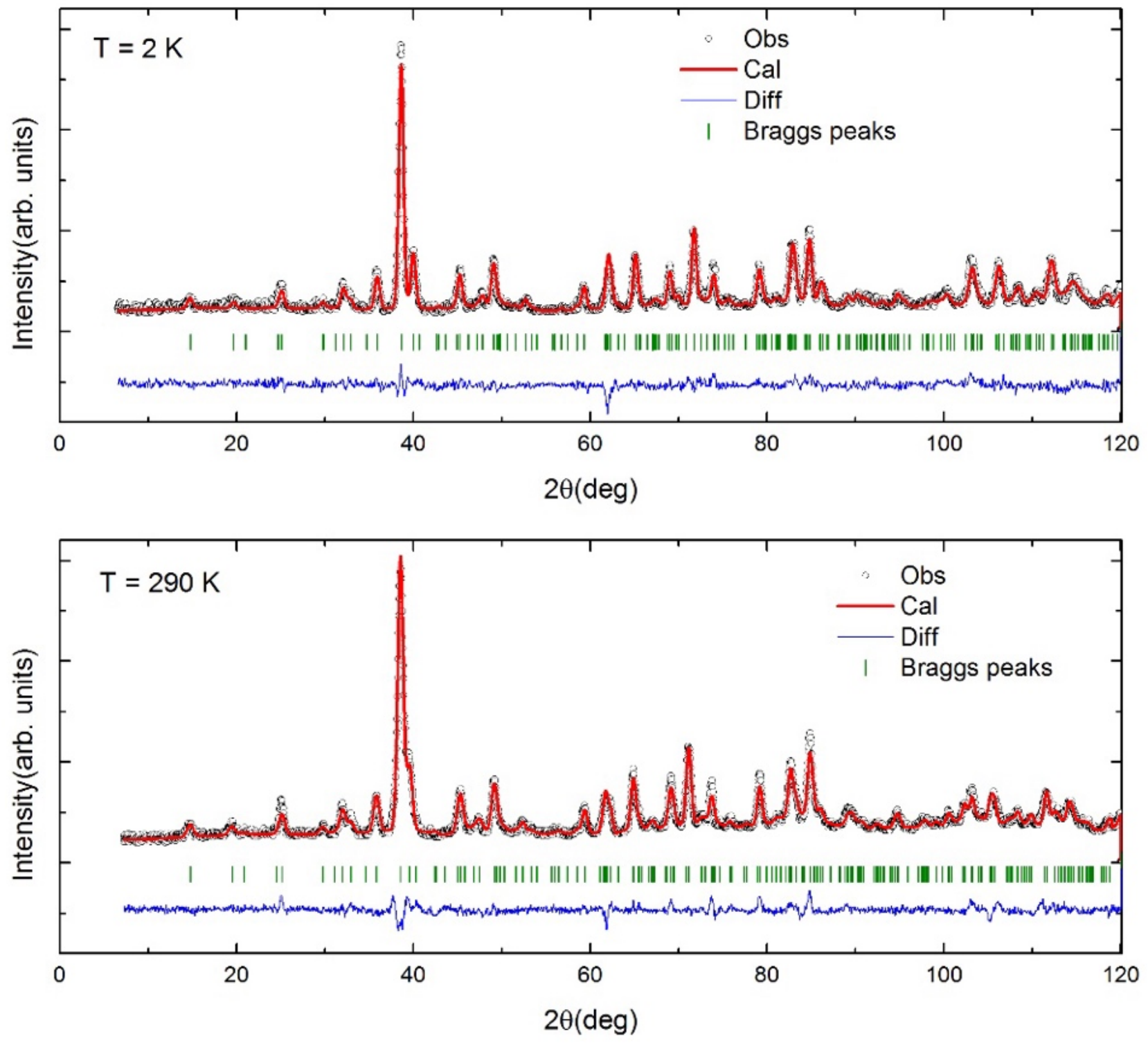

**Figure 8.** Powder neutron diffraction data for $V_2O_5$ nanoparticles were obtained at (a) 290 K and (b) 2 K respectively with no additional reflections seen at 2 K corresponding to ordering other than structural.

To further confirm the nature of magnetic order, neutron diffraction was performed on $V_2O_5$ powder samples at 290 and 2 K to compare. The same is shown in Fig 8 above at 290 K and 2 K. It was fitted with space group Pmn21 (s. g. No. 31). Table S2 in supplementary information shows the refined data from 290 and 2 K neutron diffraction, the atomic position, goodness of fitting parameters and cell parameter are also depicted in the table. Since vanadium is an incoherent scatterer, the signal-to-noise ratio is a bit noisy, which is reflected in a slightly higher $\chi^2$ value. As neutron scattering length of the element is one of the factor in structure factor calculation, which in turn help in identifying the concentration of light elements in the compound. By utilizing this property of the neutron, we observed that by introducing the deficiency in oxygen, the goodness of the fitting parameter improved significantly and we could find the chemical formula comes out to be around $V_2O_{4.886}$.

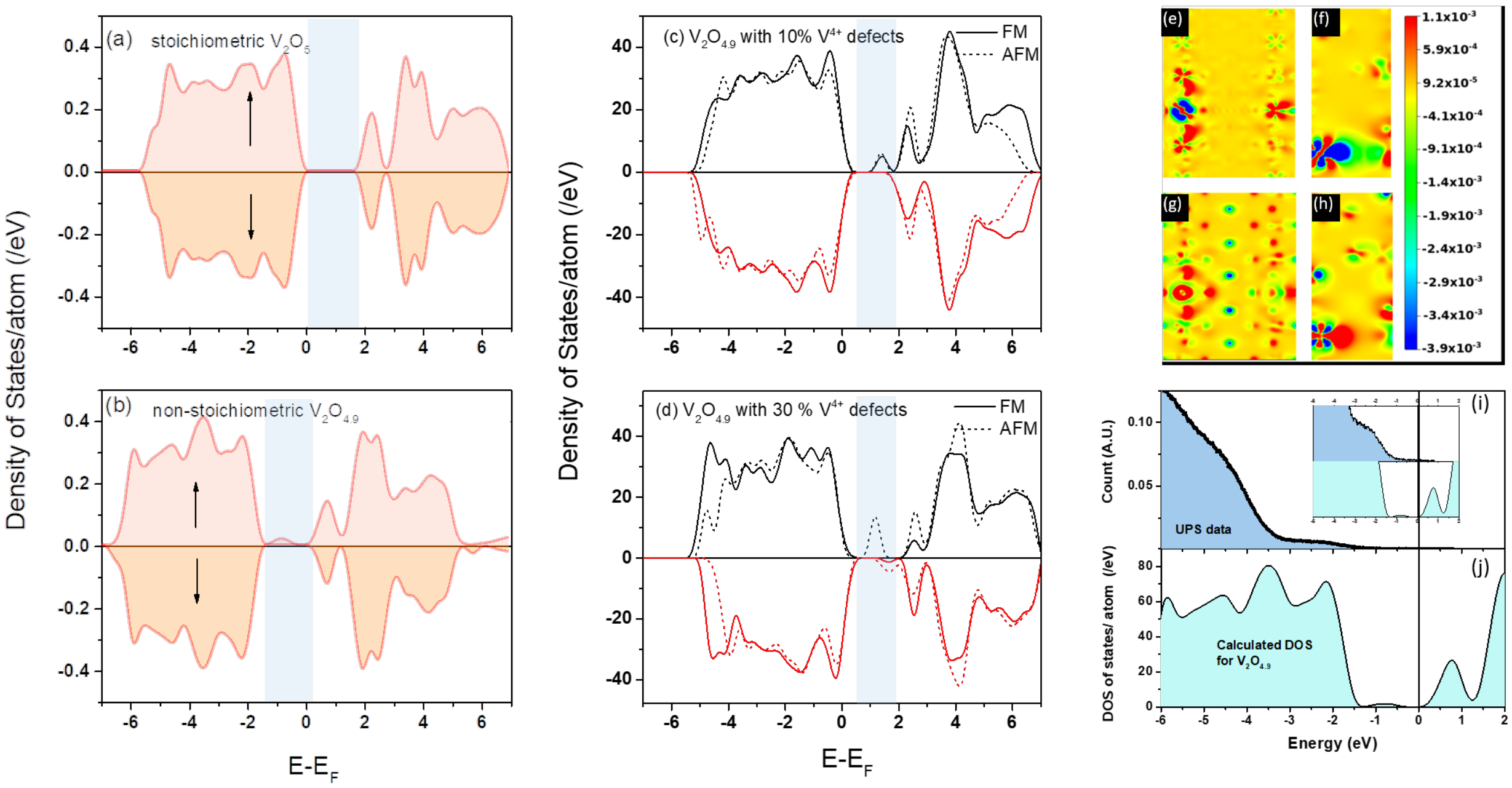

**Figure 9.** (a)-(c) Total density of states calculated with PBE functional for the cases of pure (a) $V_2O_5$, (b) $V_2O_{4.9}$ i.e. with O-vacancy. The comparison of DOS of oxygen-deficient $V_2O_{4.9}$ with (c) 10% and (d) 30 % charged defects having the FM and AFM configurations of the spins. Charge density difference plot between magnetic and non-magnetic state for ( e & f) $V_2O_5$ with O-vacancy, and (g & h) $V_2O_{4.9}$ crystal with oxygen defects as well as $V^{4+}$ ions for two planes containing the vacancy site. (Colour bar: shows positive charge density difference; higher charge density for magnetic state by red colour, while negative charge density difference higher charge density for nonmagnetic state is indicated by the rest of the colours. The comparison of (i) measured and (j) calculated DOS for valence band region with that of oxygen-deficient $V_2O_5$.

By comparing raw data as well as refined data, it is very evident that the appearance of a new peak in the pattern down to 2 K, could not be ascertained. We also did not observe an enhancement in any preexisting peak, the signature of magnetism (antiferromagnetic or ferromagnetic) could not be seen. The splitting at the 39.8$^o$ peak is more pronounced due to shrinking in the cell which has led to shifting of uneven $2\theta$ value thus improving the resolution in the peak.

### *3.4. Predictions from first principle calculations*

To estimate the origin of different magnetic contributions, the first principle calculations are performed using a supercell of vanadium pentoxide lattice stoichiometric $V_2O_5$. Further, an oxygen vacancy is created in this supercell. The stoichiometric $V_2O_5$ is known to be a diamagnetic semiconductor [76, 77], which is also verified in our study from electronic and magnetic calculations. Spin-polarised calculations result in zero value of magnetic moment on all the atoms indicating the diamagnetic state of $V_2O_5$. The total density of states (DOS) plot as shown in Fig. 9(a) exhibits semiconducting properties with a band gap of ~ 1.9 eV between the valence and conduction band calculated using PBE functional (shown in blue band). The HSE06 functions were also performed and those resulted in overestimation of the band gap to 3.6 eV. (see Fig. S8 in Supplementary Materials section). Nevertheless, the PBE results are in good agreement with those reported earlier.[78] Here, PBE stands for an exchange-correlation functional, while HSE06 stands for a hybrid functional classes of DFT (using GGA).

**Table 3.** (Colour online) Total magnetic moment ($\mu_B$/vacancy) and energy difference $\Delta E$ (eV/vacancy) between magnetic and non-magnetic states for different cases of $V_2O_5$ considered.

| | Total magnetic moment per vacancy ($\mu_B$/vacancy) | ΔE (eV/vacancy) |
|---|---|---|
| Pure $V_2O_5$ | 0 | - |
| $V_2O_5$ with O-vacancy | 1.775 | -0.788 (FM) |
| O-deficient $V_2O_5$ with 10% charged defect | 0.818 | -0.382 (AFM) **-0.383(FM)** |
| O-deficient $V_2O_5$ with 30% charged defect | 3.959 | **-3.277(AFM)** -3.273(FM) |

Electronic and magnetic properties calculations were done of $V_2O_5$ by introducing 10% oxygen vacancies in the perfect crystal structure. Spin-polarised calculations indicate a magnetic moment induced on a vanadium atom bonded to the site of oxygen vacancy as well as a small magnetic moment induced on oxygen atoms nearby to the vacancy site giving a net magnetic moment of 1.775 $\mu_B$ per unit cell per vacancy. Total energy calculations to determine ground-state magnetic ordering stabilise the ferromagnetic solution. The energy difference ΔE (in eV) between magnetic and non-magnetic states as well as total magnetic moment (in $\mu_B$) values per vacancy are listed in Table 3. The electronic spin-polarized DOS plot in the ferromagnetic state shown in Fig. 9(b) with PBE functionals, clearly indicates that the band gap value for $V_2O_5$ with O-vacancy estimated at 1.9 eV is closer to the experimentally determined value of 2.2 eV. In the case of HSE calculations, the value obtained is also large (greater than 3 eV).

The electronic and magnetic properties for the case of O-deficient $V_2O_5$ in which 10% and 30% fraction of $V^{5+}$ ions have been replaced with $V^{4+}$ ions (as charge compensatory defect) were investigated. The spin-polarised calculations stipulate that the induced magnetic moment on the V atom bonded to the vacancy site due to oxygen vacancies is reduced when $V^{5+}$ ions are replaced with $V^{4+}$ ions. Also, moments on V atoms align antiferromagnetically as well as small magnetic moment induced on O atoms as depicted by the total magnetic moment value 0.818 and 3.959 $\mu_B$ for 10% and 30% fraction of $V^{5+}$ ions respectively as shown in Table 3. Further, as shown in Fig. 9(c) and (d), the estimated DOS for the case of O-deficient $V_2O_5$ with 10% and 30% charged defect reveals metallic behaviour with fermi level trapped inside the valence band. HSE calculations yield similar behaviour to the case of PBE calculations. There is a very small change in FM and AFM DOS for both charged defect cases due to the robust local magnetic configuration of the V atom bonded to the site of O vacancy with the magnetic configuration of the system [79]. However, It should be noted that the DOS of charged defect may be less accurate due to non-localised extra electron in the system. We also performed spin-polarised calculations for the 10% charged defect in O-deficient $V_2O_5$, in which case, the ferromagnetic ordering was found to have the lowest energy. The corresponding energy values are listed in Table 3. Adding charged defects only helps in explaining AFM contribution i.e. AFM increases as $V^{4+}$ goes from 10% to 30 %. The moment value obtained for only the oxygen-deficient case is 1.77 $\mu_B$ which is close to a single electron (S=1/2) as estimated from the Curie constant value.

The charge density difference plots obtained from theoretical calculations of magnetic and non-magnetic states for $V_2O_5$ with O-vacancies are shown in Fig. 9(e) and (f). Similar plots for the $V_2O_{4.9}$ crystal with O-vacancies as well as $V^{4+}$ ions, for two planes containing the vacancy sites are shown in Fig. 9(g) and (h). The calculated DOS using PBS for crystal having oxygen vacancies has been compared with that of UV photoelectron spectrum (UPS) which gives local DOS below the Fermi level. It may be observed from Fig. 9(i and j), that there is a very good correlation observed in that of calculated and experimental spectra which denotes the experimental validity of results. It is clear that despite the presence of $V^{4+}$ surface defects, the electronic band structure does not concur with its simulated band structure for measured $V^{4+}$ defect densities. Hence the contribution of such $V^{4+}$ defect is minimal in deciding the electronic energies. Besides, the amount of charged defect contributing (10 or 30 %) cannot be large here as by finite temperatures the thermal activation helps bound polarons to become free, resulting in more localised electrons[80]. It has also been seen from the transport measurements that temperature dependence of conductivity becomes band-like (1/T). Whereas, the thermopower measurements and Heike's formula tell us that the effective $V^{4+}/V^{5+}$ ratio should be one in a thousand or lower. (See supplementary material Fig. S2). Thus, only a small fraction of charged defects leads to polaronic states. Others may contribute to the paramagnetic background. Therefore, none of those DOS for charged defects matched the observed UPS. The electronic band structure is governed by metal $d$ and O $2p$ characters of the bands and the mid-gap states arise as a result of oxygen deficiencies. These mid-gap states are characteristics of polarons in $V_2O_5$ (like other polaronic solids) [80].

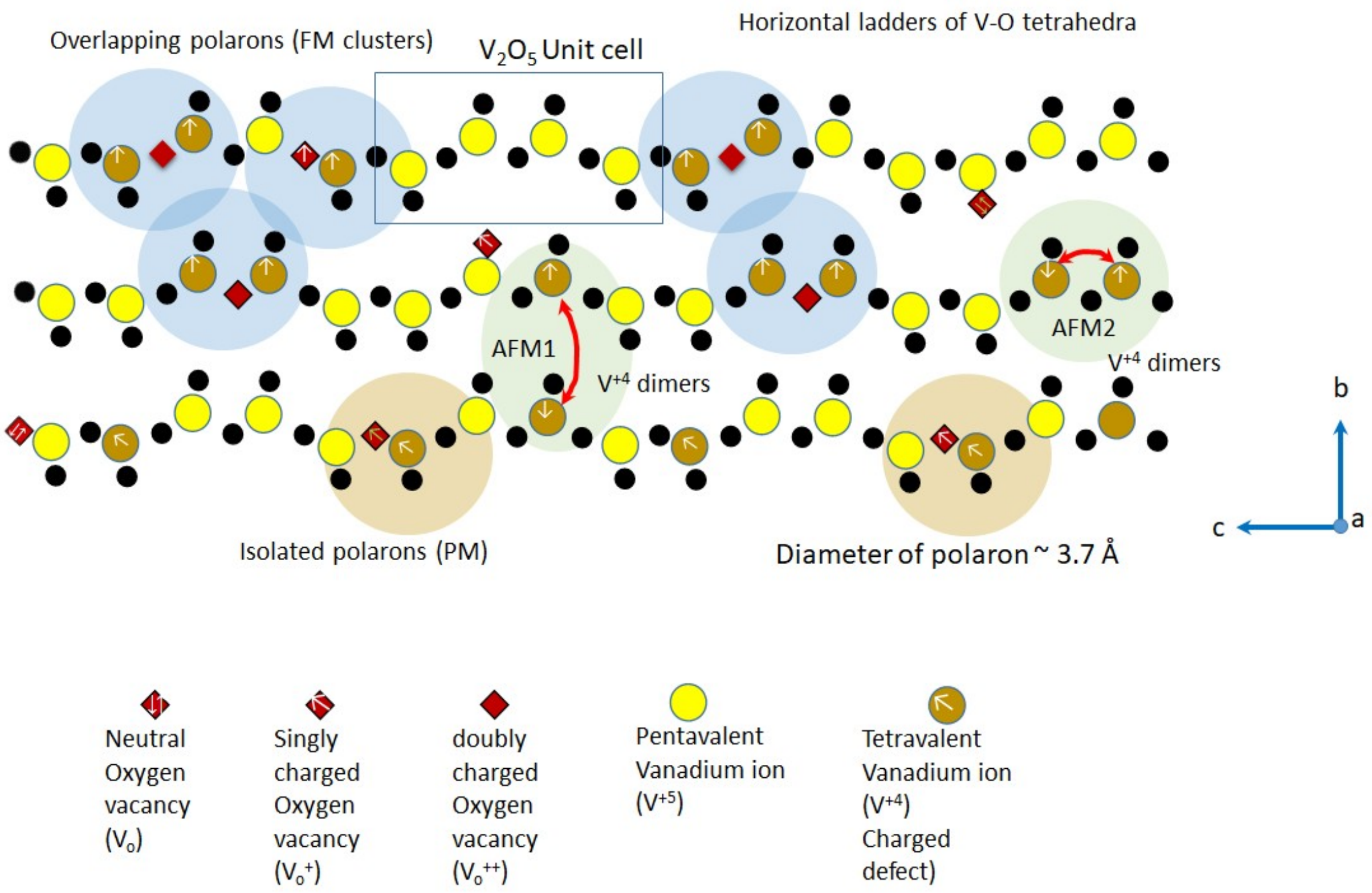

**Figure 10.** The schematic diagram shows the various possible magnetic interactions in an undoped $V_2O_5$ lattice having oxygen vacancies (charged or neutral) and compensating $V^{4+}$ defects, together having a bound state called polaron. These can be isolated or overlapping polarons depending on inter-polaron distance and hence may result in paramagnetic or ferromagnetic interaction. The pair of $V^{4+}$ ions (dimers) could also couple antiferromagnetically within the same or neighbouring strands of the ladder. Please note that the figure is merely for the better explanation of different kinds of defects that may occur in $V_2O_5$ lattice and their interactions which may lead to polarons as well as magnetic exchanges of various kinds. Due to limitations of space and better clarity, all these have been drawn in a very small space of $V_2O_5$ lattice. This should not be confused with the actual defect densities which exist in the experiment, nor those used in simulations for modelling the system.

## 4. Discussion

Although bulk $V_2O_5$ is diamagnetic, experimental magnetic measurements conducted on the oxygen-deficient nanoparticles showed a large variation in the observations ranging from ferromagnetism[28, 30, 33], antiferromagnetism[69] and to a mixture of both[29, 31]. While first principles calculations performed by Xiao et.al.[32] predicted that the $V_2O_5$ system may have vacancy-induced ferromagnetism if the x-value lies within 0.19 to 0.45 or below 0.13. However, through $0 < x < 0.5$, it shall also exhibit an antiferromagnetic coupling (AFM-1 shown in Fig 10) because of antiparallel coupling of $V^{4+}$ spins on adjacent ladders, (often referred to as dimers[29]). Besides, at the $x > 0.25$, another AFM coupling (AFM-2 shown in Fig 10) may arise due to anti-parallel coupling of $V^{4+}$ spins on adjacent rungs of the ladder. For low vacancy concentrations (and low $V^{4+}$ defects) the electronic structure remains similar to that of bulk $V_2O_{5-x}$, while for large oxygen vacancy concentrations, the electronic structure may exhibit a significant change as also confirmed by our calculations. Thus, we envisage that the electronic structure of our sample is likely to be similar to that of oxygen-deficient $V_2O_{5-x}$ with a trace amount of $V^{4+}$ defects present. This can be readily confirmed upon comparison of the UPS spectrum of our sample with the calculated DOS in Fig. 9(i-j), which shows very good agreement. Besides, the $x$ value for our sample (~0.35), as calculated from the XPS data, legitimises the presence of a combination of both FM and AFM phases in it (both AFM-1 as well as AFM-2). This is seen as the AFM to PM transition at about 280 K in the Curie Weiss plot shown in Fig. 4(d). Here $V^{4+}$ introduces the AFM interaction and increases with increasing its amount as seen from the calculations.

Bhandari et.al. [81] investigated the effect of electron doping on the electronic and magnetic properties of $V_2O_5$. Here they employed several strategies to dope the system with electrons, such as varying the effective atomic numbers and intercalation of alkali metal within the layers of $V_2O_5$. Similar electron doping occurs when the system is doped with oxygen vacancies such as in this case. Wherein the generalised Heisenberg Hamiltonian used is of the form,

$$H = \sum_{i \neq j} J_{ij} \boldsymbol{e_i} . \boldsymbol{e_j} \quad (7)$$

Counting the spin on each non-equivalent vanadium atom of the two parallel rungs twice. The interactions $J_1$, $J_2$, $J_3$ were identified among the nearest neighbour vanadium ions and the net spin was found to be 0.5 $\mu_B$ on each vanadium ion due to the excess of electrons distributed over all the vanadium atoms effectively rather than electron localisation on the defect. The energy difference calculated between FM and AFM shows that below a particular electron concentration of 0.88 electrons per vanadium atom, strong FM is shown by the sample, while on exceeding this it tends to show AFM dominance. The maximum exchange coupling of $J_1$ ~ 44.8 meV is observed between two parallel rugs of the ladder with an electron delocalised over each of them. i.e. same electron is shared by two vanadium atoms of the same rug of the latter (making unit cell) (See Fig 10) and hence having the same spin. Whereas the j-values of the other two nonequivalent sites predicted are $J_2$ = -4.5 meV and $J_3$ = -2.1 meV. Thus, quoting from Bhandari et. al. [81] The nature of both interactions is antiferromagnetic. They reduce significantly as distance increases. This is an indication of neighbouring chains coupling antiferromagnetically and the within chain ordering is also antiferromagnetic. This has been marked in Fig 10 as AFM1 and AFM2.

It is perhaps wiser to examine the polaronic signature observed in the electrical transport studies and the concurrently observed magnetic behaviour, due to their common origin from an oxygen vacancy paired with a $V^{4+}$ defect. An oxygen vacancy can be neutral ($V_O$) singly charged ($V_o^{+}$) or doubly charged ($V_o^{++}$) making shallow and deep donors. Most of the existing literature has overlooked or only marginally discussed this under the umbrella of bound magnetic polarons (BMP), without any reference to their electrical contributions. The problem of polaronic conduction and its energetics has been a question of interest for a long time. Sanchez et.al.[61] in their paper published in 1982 showed a remarkable correlation between DC electrical conductivity and Electron Spin Resonance (ESR) experiments, to establish the mechanism of polaronic conduction in the $V_2O_5$ system. The study revealed that there exists a coupling between the unpaired electron of $V^{4+}$ and an associated oxygen defect. The electron may be localized on equivalent $V^{5+}$ ions in the vicinity of the defect (as shown by the blue/brown shaded circle in Fig 10). Nevertheless, this polaron is a small polaron which could be localised near the defect (i.e. bound polaron) or it could be localised (free polaron) when thermally activated with a sufficient activation energy of about 0.2 eV, leading to increased conductivity. Thus, the electrical conductivity quickly escalates as the polaron hopping is enabled with sufficient thermal activation. The activation energies mentioned by Sanchez et.al.[61] are in very good agreement with those calculated in this study.

We estimated stoichiometry from neutron diffraction via refinement. The modelled composition of $V_2O_{4.886}$ is also very close to $V_2O_{4.9}$ used in DFT modelling whose band structure matched with that of DOS estimated using UPS data. This system can be considered as equivalent to a $d^0$ semiconductor, self-doped with trace transition metal ($V^{4+}$) impurities similar to those described in reference [23]. As mentioned earlier, at low temperatures, the polarons are bound and the unpaired electrons are localised near the defects. These being small polarons, they are reasonably near, yet isolated from each other. The calculated radius of a small polaron (1.84 Å) is a good indicator of this. Isolated polarons are distributed throughout space and can contribute to the paramagnetic background, which decreases sharply with temperature. The bounded electrons, confined in polaronic states in their vicinity, may show a direct ferromagnetic exchange, referred to as polaron-polaron interaction. However, their separation distances may play a key role in deciding the dominant interactions (See Fig 10). The background semiconductor matrix may meanwhile have a carrier-carrier antiferromagnetic interaction from ions.

As predicted by Durst et.al.[23], the susceptibility variation shown in Fig. S8, in the temperature of interest, shows a dual magnetisation mechanism i.e. low field involving polaron-polaron interaction which could induce direct ferromagnetic exchange. Whereas high field isolated polarons and among the magnetic ions, if any. While the linear slope at a high field (> 1000 Oe) results from paramagnetic spin alignments.

Owing to the nano-sized nature of the particles additional effects may be expected, such as superparamagnetism from the core of particles (domains), as observed here. Besides, the intergranular dipoles may contribute to the paramagnetic background. Further, due to the orientational anisotropy of the particles, the relative orientation of the easy axis with the field shall have a precision contributing to spin-flip distribution (abrupt along the easy axis and gradual perpendicular to the easy axis).

Subsequently, at high temperatures, as the polarons become free by thermal activation, long-range correlations are enabled. These polaron-polaron correlations observed at low temperatures and low fields are disturbed, resulting in gradually vanishing magnetism at high temperatures. Nevertheless, at high temperatures, a weak ferromagnetic hysteresis at 300 K, suggests the existence of short-range FM clusters, which could be primarily due to the dilute magnetism.

In our other works[49], it has been shown that doping phosphorus at the vanadium site leads to a reduction in the hysteresis loop and introduces a positive slope in M-H at high fields, implying dominating PM behaviour. Besides, in the same work, it is demonstrated that the PM nature is manifested in higher M-H slopes and sharper EPR signals with increasing P doping.

Thus, we have a tentative picture of the competing processes that are present in the system and how they contribute to the magnetic data. Fig. 10 is a pictorial representation of that. There are $V^{4+}$ ions besides $V^{5+}$ ions in the material. The O-vacancies capture electrons in them and they hybridise with the d-orbital of the nearby $V^{4+}$ ions to form polarons. Polarons can be thought of as large bubbles containing spins, spanning several formula units (FU). Hence isolated polarons will have a paramagnetic moment as seen in EPR spectra. When they come close enough, they can overlap and give rise to FM or AFM in clusters since no long-range order is observed. As in the sample, all these polarons and $V^{4+}$ ions are distributed at random, and depending on the spacing between them, RKKY interaction decides if the coupling will be FM or AFM. Hence there is an inherent randomness and competing interactions within the system and it is conducive to spin-glass behaviour, which might be the reason for the split in the FC-ZFC curves. For confirmation, AC susceptibility measurements are required.

## 5. Conclusion

The study of the magnetic properties of otherwise bulk diamagnetic vanadium pentoxide reveals the significance of oxygen vacancies and vanadium ion charge stoichiometry in imparting magnetic moment to the sample. The electrical transport studies reveal a polaronic hopping mechanism due to the presence of oxygen vacancies and charged vanadium defects. Electrical resistivity and Seebeck coefficient show the polaron hopping mechanism with a hopping energy of 0.112 eV. The radius of the polarons is estimated to be 1.8 Å. Polaron binding energy of 0.287 eV confirms the possibility of long-range hopping at sufficiently high temperatures causing a transition from bound to free polarons, thereby improving the electrical conductivity at higher temperatures. Electronic and magnetic calculations for three cases of pure $V_2O_5$, $V_2O_5$ with 10% O-vacancy, and O-deficient $V_2O_5$ with 10 and 30% $V^{5+}$ ions replaced with $V^{4+}$ ions have been studied theoretically. Whereas pure $V_2O_5$ is a diamagnetic semiconductor, $V_2O_5$ with 10% O-vacancy is a ferromagnetic semiconductor. On introducing charged defects, $V_2O_5$ with 10% O-vacancy is found to be an antiferromagnetic metal. Thus, it is proposed that vanadium-charged defects are in dilute limit which induced antiferromagnetism within dimers but no changes in the band structure. Experimental observations are in good agreement with the predictions of first-principles calculations. The findings are concurred with techniques like EPR and neutron diffraction at different temperatures. It shows a ferromagnetic ordering for low temperatures which is attributed to polaron-polaron interactions at low fields and isolated spin alignments at high fields. Nevertheless, there also exists antiferromagnetic contributions to the magnetic signals, due to vanadium dimers across the ladders in the crystal structure of $V_2O_5$. For that further analysis and time-dependent measurements are required, like AC susceptibility.

**Acknowledgements**

The authors are grateful to the institute SQUID magnetometer facility of IISER Thiruvananthapuram for the magnetic measurements Dr. Deepshikha Jaiswal-Nagar, Dr Tuhin S. Maity, Mr Waseem for their help and Support during the magnetic measurements and Hall effect Measurement systems (HEMS) for resistivity, mobility and carrier concentration measurement at low temperatures. The funding received from the Science and Engineering Board (SERB), Government of India for the research grant (EEQ/2018/000769, EEQ/2022/001016) supporting this work, is gratefully acknowledged. The authors express their gratitude to Mr. Akshay S. Kamble from UGC CSR Mumbai for his help in the refinement of neutron diffraction data.

**Data availability statement**

The data that support the findings of this study are available upon request from the authors.

Supplemental Material

# Coexistence of Multiple Magnetic Interactions in Oxygen Deficient $V_2O_5$ Nanoparticles

Tathagata Sarkar[1], Soumya Biswas[1], Sonali Kakkar[2], A. V. Raghu[1], S. D. Kaushik[3], Chandan Bera[2], Vinayak B. Kamble*

[1]School of physics, Indian Institute of Science Education and Research Thiruvananthapuram, India 695551.

[2]Institute of Nano Science and Technology, Habitat Center, Phase-X, Mohali, Punjab 160062, India.

[3]UGC – DAE Consortium for Scientific Research Mumbai Centre, Bhabha Atomic Research Centre, Mumbai-400085, India.

## A. The Rietveld refinement parameters

Table S1. The summarised list of X-ray Rietveld refinement parameters.

| Cell parameters (Å) | a (Å) | b (Å) | c (Å) | Vol (Å³) | α | β | γ |
|---|---|---|---|---|---|---|---|
| | 11.5110 (3) | 4.3739(1) | 3.5640(4) | 179.440(2) | 90 | 90 | 90 |
| Space group | P m n 21 | | | | | | |
| **Atom** | **Type** | **X** | **Y** | **Z** | **Occupation** | | |
| V1 | V | 0.1481(2) | 0.1025(4) | 0 | 1 | | |
| O1 | O | 0 | 0.0175(1) | -0.0096(4) | 0.5 | | |
| O2 | O | 0.1441(3) | 0.4530(2) | 0.9976(3) | 1 | | |
| O3 | O | 0.1817(5) | 0.0070(2) | 0.4935(8) | 1 | | |
| | | | | | | | |
| **χ2** | | 3.653 | | | | | |

Table S2. The summarized list of neutron Rietveld refinement parameters at 2K and 290 K.

| Cell parameters | a (Å) | b (Å) | c (Å) | Vol (Å³) | α | β | γ |
|---|---|---|---|---|---|---|---|
| **(A) at 290 K** | 11.5252(5) | 4.3706(7) | 3.5603(6) | 179.342(4) | 90 | 90 | 90 |
| **(B) 2 K** | 11.5245(3) | 4.3287(4) | 3.5593(2) | 177.527(7) | 90 | 90 | 90 |
| Space group | | P m n 21 | | | | | |
| **Atom** | **Type** | **X** | **Y** | **Z** | **Occupation** | | |
| V1 | V | 0.1667(4) | 0.06231(5) | 0.00000 | 0.951 | | |
| O1 | O | 0 | 0.0051(3) | -0.08508(3) | 0.436 | | |
| O2 | O | 0.1458(3) | 0.4703(9) | 0.8995(2) | 0.940 | | |
| O3 | O | 0.1798(1) | 0.0031(5) | 0.4063(4) | 0.946 | | |
| | | | | | | | |
| | | 290 K | | 2 K | | | |
| **χ2** | | 11.3 | | 15.2 | | | |
| **Rp** | | 5.68 | | 7.27 | | | |
| **Rwp** | | 7.25 | | 9.41 | | | |
| **Rex** | | 6.36 | | 6.47 | | | |

| Atom | Type | X | Y | Z | Occupation |
|---|---|---|---|---|---|
| V1 | V | 0.1667(4) | 0.0623(5) | 0.0000 | 0.951 |
| O1 | O | 0 | 0.0050(3) | - 0.0859(3) | 0.436 |
| O2 | O | 0.1458(2) | 0.4703(9) | 0.8992(2) | 0.940 |
| O3 | O | 0.1798(1) | 0.0031(5) | 0.4063(4) | 0.946 |

## B. Reflectance Measurement and Bandgap estimation

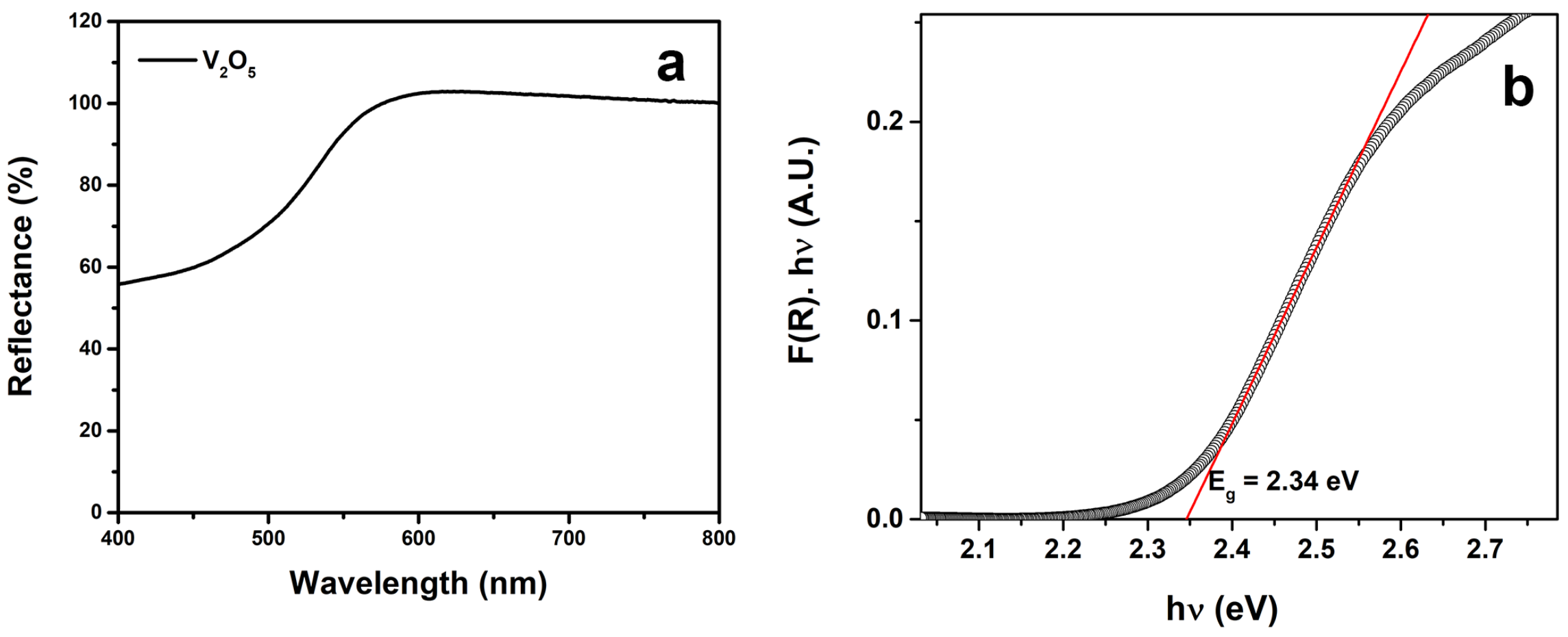

**Figure S1**. (a) The reflectance spectra of the sample and (b) Tauc plot using KM function.

The reflectance of the drop casted sample was measured. It is shown in Fig. S2 (a). The reflectance is nearly 100% for longer wavelengths and drops sharply for wavelengths below 550 nm. The reflectance was used to calculate the absorption coefficient α using the Kubelka Munk function (F(R)) shown below

$$F(R) = \frac{\alpha h\upsilon}{s} = \frac{1-R(\lambda)^2}{2R\ (\lambda)} \qquad \text{(S1)}$$

where, s is the scattering factor. Thus, F(R) can be replaced with α to obtain bandgap value $E_g$ in the following equation

$$\alpha h\upsilon = k(h\upsilon - \mathrm{Eg})^{1/n} \qquad \text{(S2)}$$

where n=1/2 for direct and n=2 for indirect transitions.

Thus, the indirect bandgap calculated from the KM plot is found to be 2.3 eV. Whereas, for the direct transitions it is 2.45 eV along with a small peak like absorbance centred at 2 eV.

## C. Thermopower temperature dependence and defects

The Heike's formula is given by[1]

$$S(T \rightarrow \infty) = \frac{k_B}{e} \ln\left(\frac{c}{1-c}\right) \quad \text{(S3)}$$

Where c is the fraction of $V^{4+}$: $V^{5+}$.

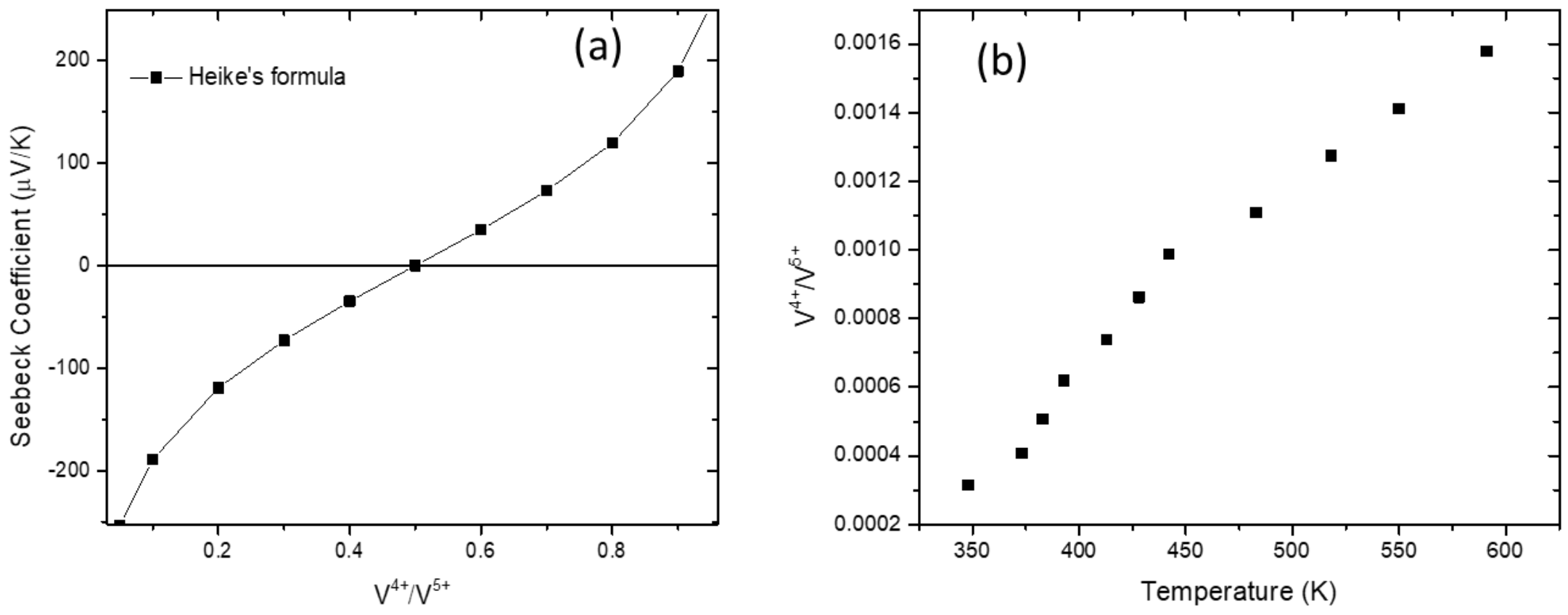

**Figure S2**. (a) The seebeck coefficient variation with $V^{4+}/V^{5+}$ and (b) observed $V^{4+}/V^{5+}$ fraction for measured Seebeck coefficient of $V_2O_5$ sample.

## D. Blank data of SQUID magnetometer.

The squid magnetometer sample holder data was measured and the same has been shown below in **Figure S3.**

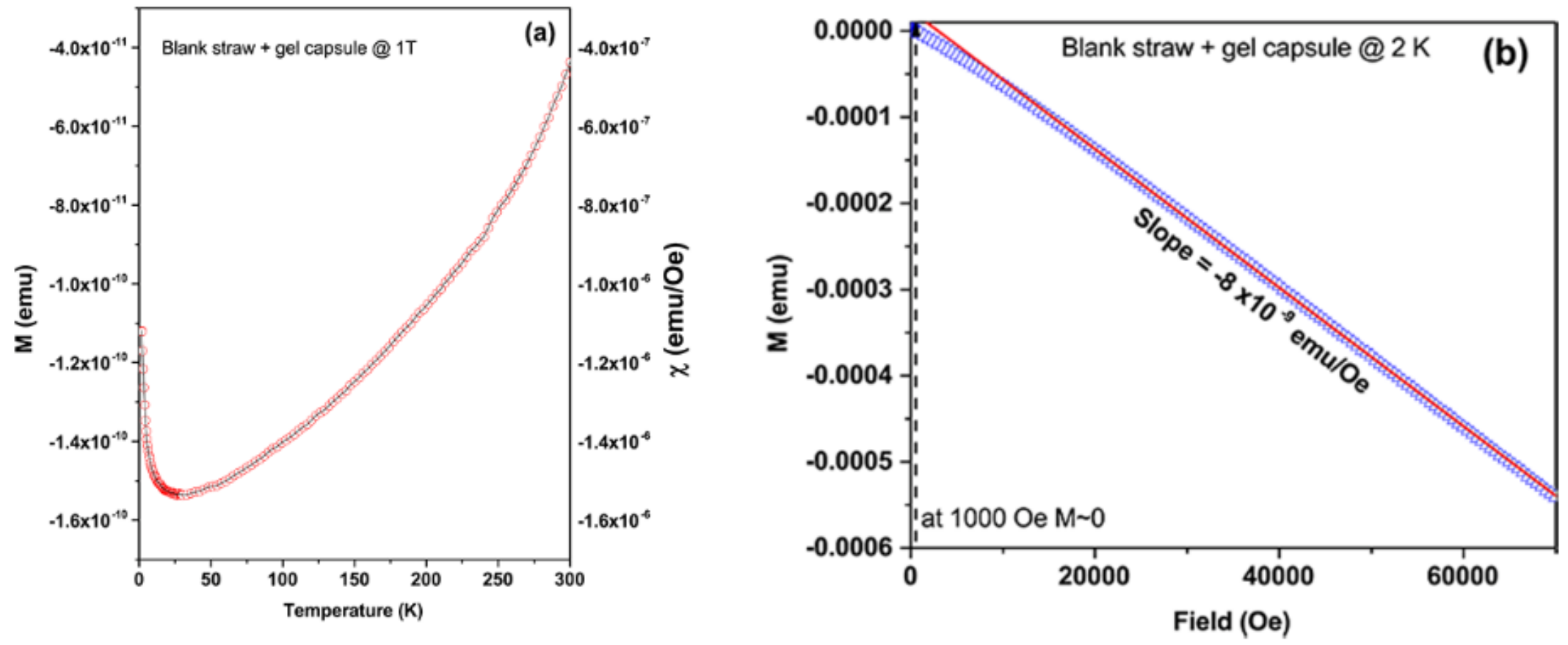

**Figure S3**. The measurement of background (blank) data of the sample holder in SQUID magnetometer (a) with temperature at 1 T field, and (b) with field at 2 K.

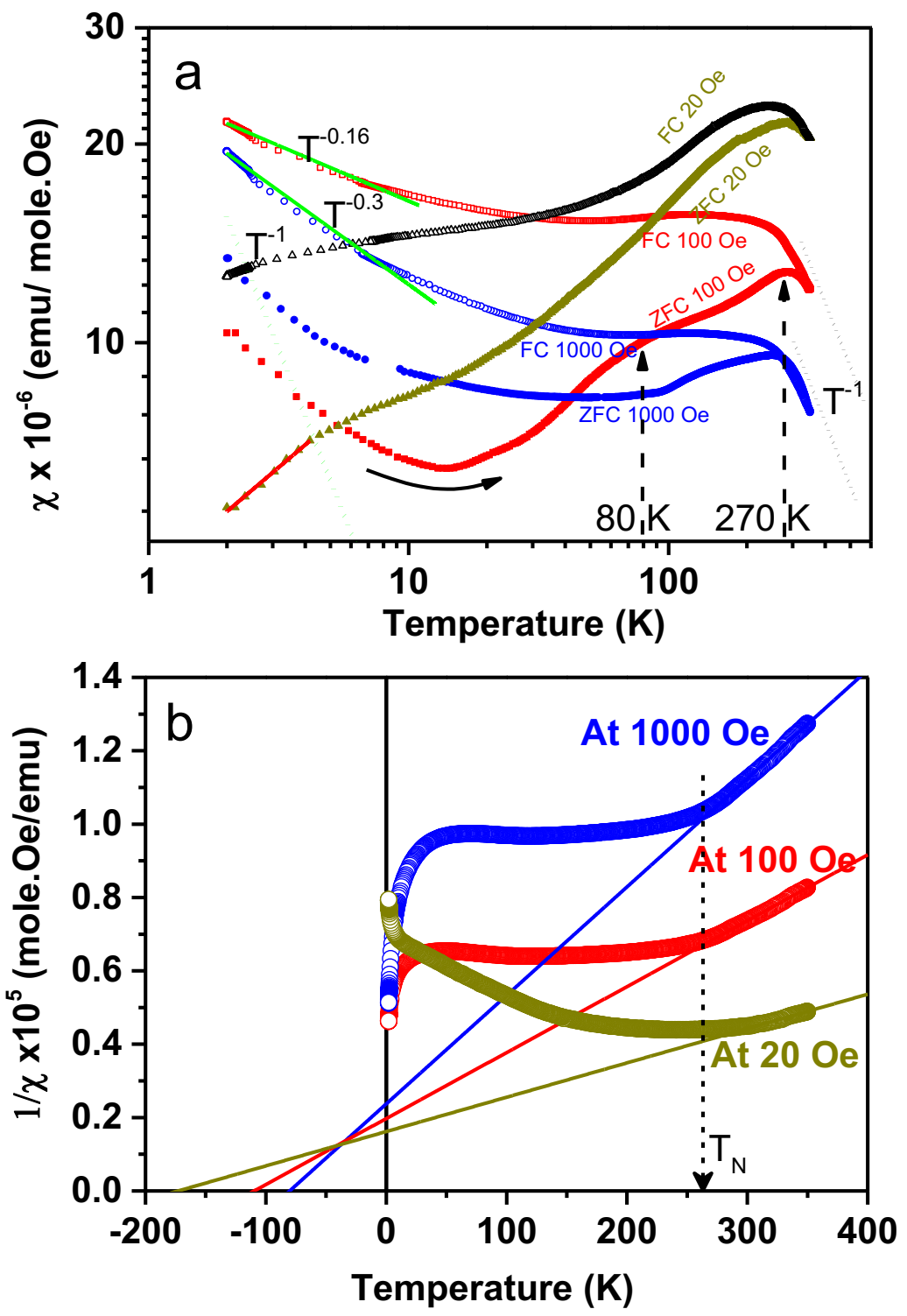

**Figure S4**. The comparison of the ac susceptibility vs T and curie Weiss plots of $V_2O_5$ at different bias fields.

## E. Likely Experimental signatures of suparparamagnetism

A clear signature of blocking temp in FC- ZFC etc and zero coercivity above this temperature should be seen if there exists any superparamagnetism as seen in case of Guria et.al. whose findings on $Sm_2O_3$ flower-morphologies is close to our results[2].

The similarities of our work with that of Guria et. al. that investigates "Tuning the growth pattern in 2D confinement regime of Sm2O3 and the emerging room temperature unusual superparamagnetism" include the following.

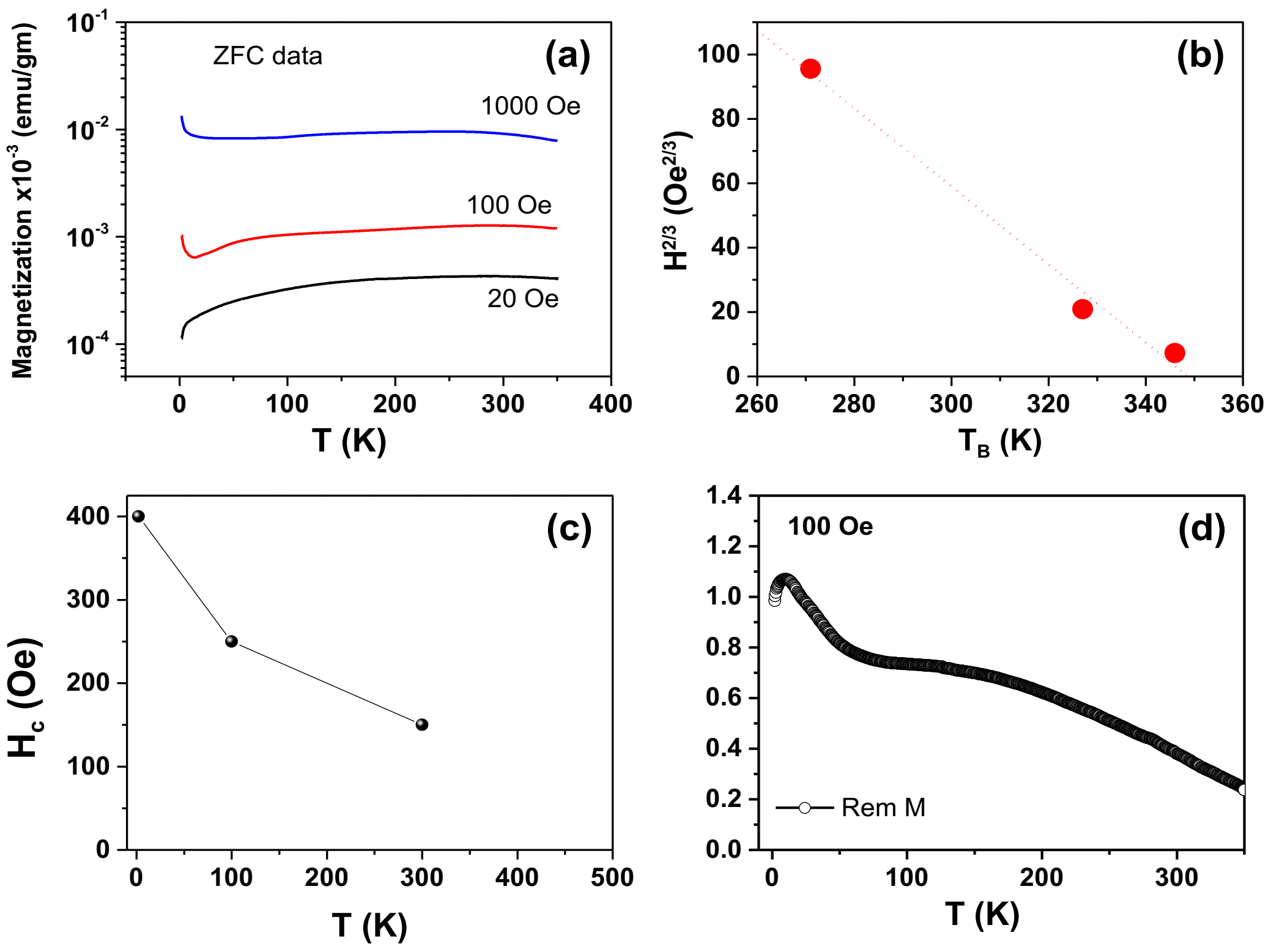

**Figure S5.** (a) The ZFC vs Temp data at different field for our measurements. (b) The Field^2/3 vs Blocking temperature of the FC-ZFC data. (c) The coercivity of the M-H data. The same may be compared with literature for experimental signature of SPM[2].

Nevertheless, in our case, the blocking or bifurcation temperature in FC-ZFC is higher than 300 K for 100 Oe, therefore we do not see the coercivity diminishing to zero in measured temperature range unlike Guria at.al.

## F. Langevin and Brillouin Function fitting to M-H data

In order to confirm the PM and FM contributions, and estimate the moment in the low temperature the M-H data is fitted with the Brillouin and Langevin function.

Equations below have been used to fit the M-H curve. It consists of two parts - one is the Langevin function and the other is a simple paramagnetic $\chi_0 H$ term. It has been used because a single Langevin function couldn't successfully fit the entire data, especially the un-saturating part of the data.

Langevin function is given by,

$$M = N\mu[coth(\mu H/k_B T) - 1/(\mu H/k_B T)\,] + \chi_0 H \tag{S4}$$

where, N is number of magnetic centers (ions or spins) per unit volume. The data is converted from emu/mole to $\mu_B$ per formula unit (FU). $\mu$ is the total moment carried by each ion/spin ($\mu_B$), $k_B$ is boltzmann constant and T is absolute temperature.

As $J \rightarrow \infty$, the Brillouin function tends to the Langevin. We attempted fitting the data with Brillouin function but it could not be fit satisfactorily. Besides, the reason why Langevin had work better than Brillouin (the large value of *J*) is at this temperature most of the particles behaves like single domain ferromagnets i.e. superparamagnetism.

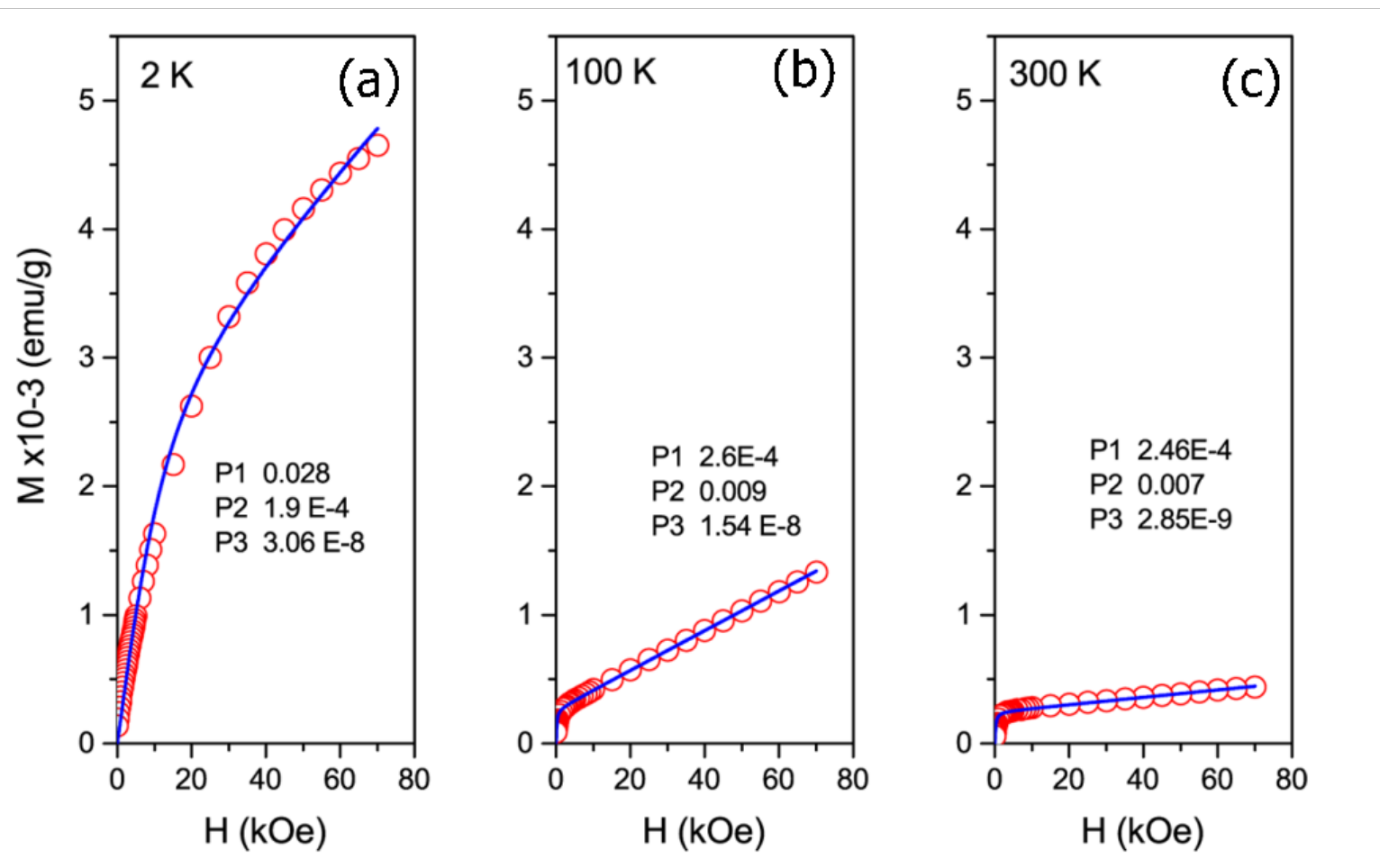

**Figure S6.** The Langevin Fitting for M-H data measured at (a) 2K, (b) 100 K and (c)300 K.

Brillouin function fitting to M-H data

The Brillouin function is given by,

$$B_J(x) = \left(\frac{2J+1}{2J}\right) coth\left(\frac{2J+1}{2J}x\right) - \left(\frac{1}{2J}\right) coth\left(\frac{1}{2J}x\right) \quad \text{(S5)}$$

Where, $x = J\frac{g\mu_B B}{k_B T}$

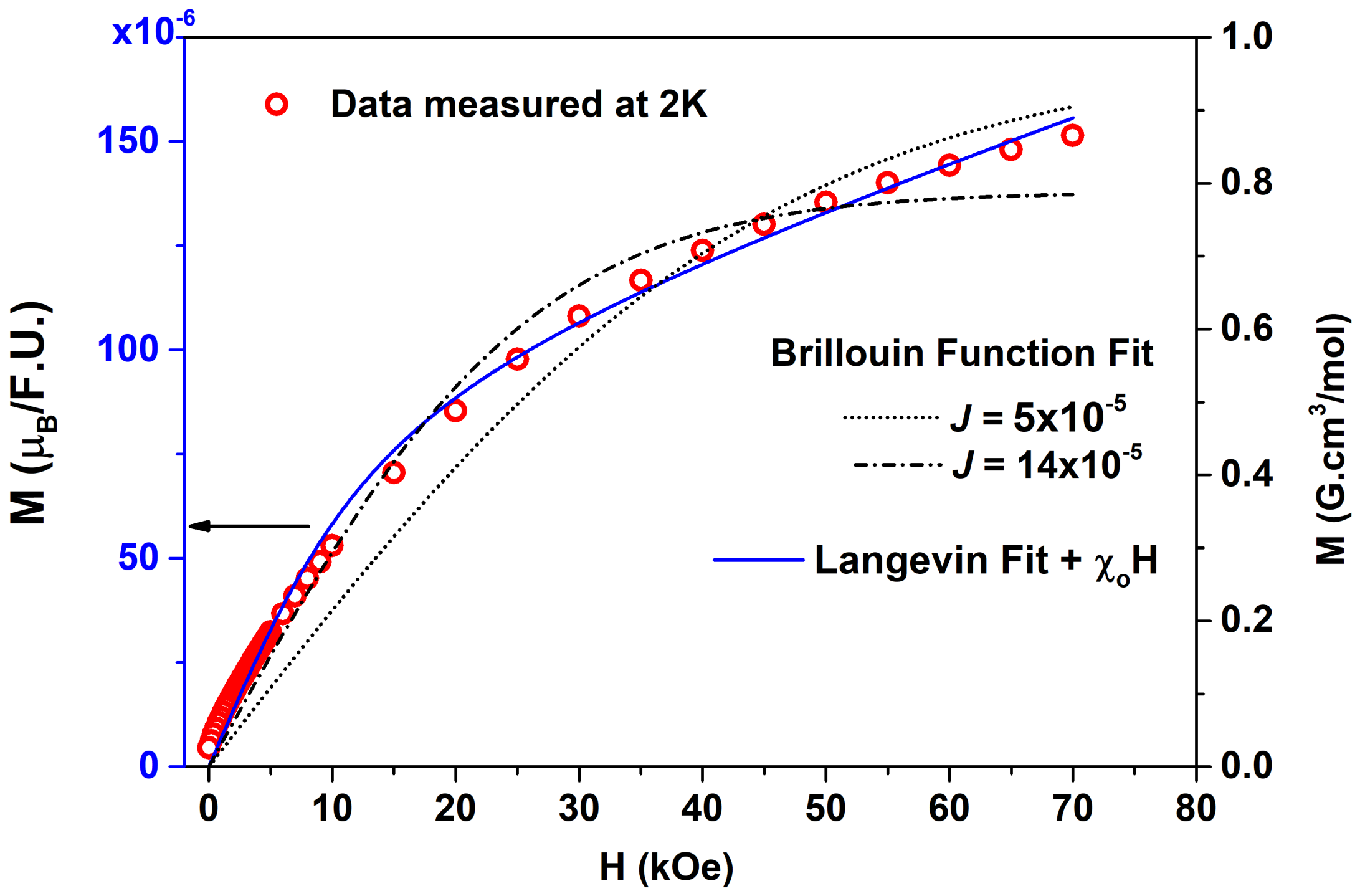

**Figure S7.** The comparison of Langevin and Brillouin function fitting to the 2K M-H data.

The attempts were made for fitting the data with Brillouin function but it could not be fit satisfactorily. The Brillouin function (shown above in eq S5 fitting for the same data did not converge and the fitting having best goodness of fit showed very low total J value (~ $10^{-5}$, see supporting information Fig S5). The results of fit can be found in supplementary material Fig S4 and S5. Although the fitting was attempted at all of the three different temperatures, only 2 K data gave physically reliable fitting parameters i.e. μ and N which are summarized in Table S1. In order to get a perfect fit, an additional $\chi_o H$ term was required along with the Langevin function Eq S4, which marks the background susceptibility.

Table S3. Results of Langevin fit with Eq. S4 to M-H data,

| T (K) | M, moment per magnetic center ($\mu_B$) | N, No of magnetic centers (per FU) | No of FU contributing single Spin |
|---|---|---|---|
| 2 | 5.65 | 1.3 x $10^{-4}$ | 7520 |
| 100 | 13.4 x$10^{3}$ | 5.22x$10^{-10}$ | 2x$10^{9}$ |
| 300 | 31.25x$10^{-3}$ | 2.1 x$10^{-10}$ | 5x $10^{9}$ |

As can be seen from Table S1, the value of $\mu$ obtained for 2 K fitting is although higher side but fairly reasonable whereas those of the high temperature are unusually large and hence unreliable (See the supporting material, Fig S6). Nevertheless, the 2 K data predicts that there is a moment of 5.6 $\mu_B$ and about 7520 formula units make up a single spin unlike traditional magnetism where half integral multiple spins are contributed by unique magnetic ions in the lattice. If this the case, this could be due to the delocalization of vacancy contributed unpaired electrons having an unpaired spin as discussed in the polaron model. Eq. S4 has been used to fit the M-H curve at 2 K. It consists of two parts - one is the Langevin function and the other is a simple paramagnetic $\chi_0$H term. It has been used because a single Langevin function couldn't successfully fit the entire data, especially the un-saturating part of the data. Besides, the reason why Langevin may have work better than Brillouin (as J → ∞ is at this temperature (2 K) most of the particles behaves like single domain ferromagnets i.e. superparamagnetism. In addition, the average distances among the magnetic spins is fairly large yielding a weak exchange (lower J). Thus, the Langevin fit worked and yielded a value of moment (5.65 $\mu_B$). Nevertheless, the number of electrons which may have contributed to give rise to such moment cannot be estimated conclusively and needs further investigations. However, the PM dominates at 2 K as is evident from the $\chi$-T curves (Fig. 5c). At low temperatures, the paramagnetic component heavily dominates over every other magnetic phase in the sample. The exact origin of each of the magnetic contributions needs to be confirmed further hence we perform first principle calculations which are discussed in the next section.

## G. Theoretical calculations comparison using HSE and PBE

We calculated the DOS of pure $V_2O_5$, $V_2O_5$ with O-vacancy, and O-deficient $V_2O_5$ with 30% charged defect using the screened HSE06 hybrid functional in order to verify the accuracy of the electronic properties calculated with PBE functional. The resulting plots with PBE and HSE functionals are displayed in Figure below.

As seen in the figure, HSE overestimates the band gap of pure $V_2O_5$ to 3.6 eV in comparison to the experimentally reported value of 2.2 - 2.3 eV. The band gap value for $V_2O_5$ with O-vacancy estimated by PBE at ~1.9 eV, as shown in fig. S6(b), is closer to the experimentally determined value of 2.2 eV. As shown in Fig. S6(c), both the HSE as well as PBE estimated DOS for the case of O-deficient $V_2O_5$ with 30% charged defect reveals metallic behaviour with fermi level trapped inside the valence band.

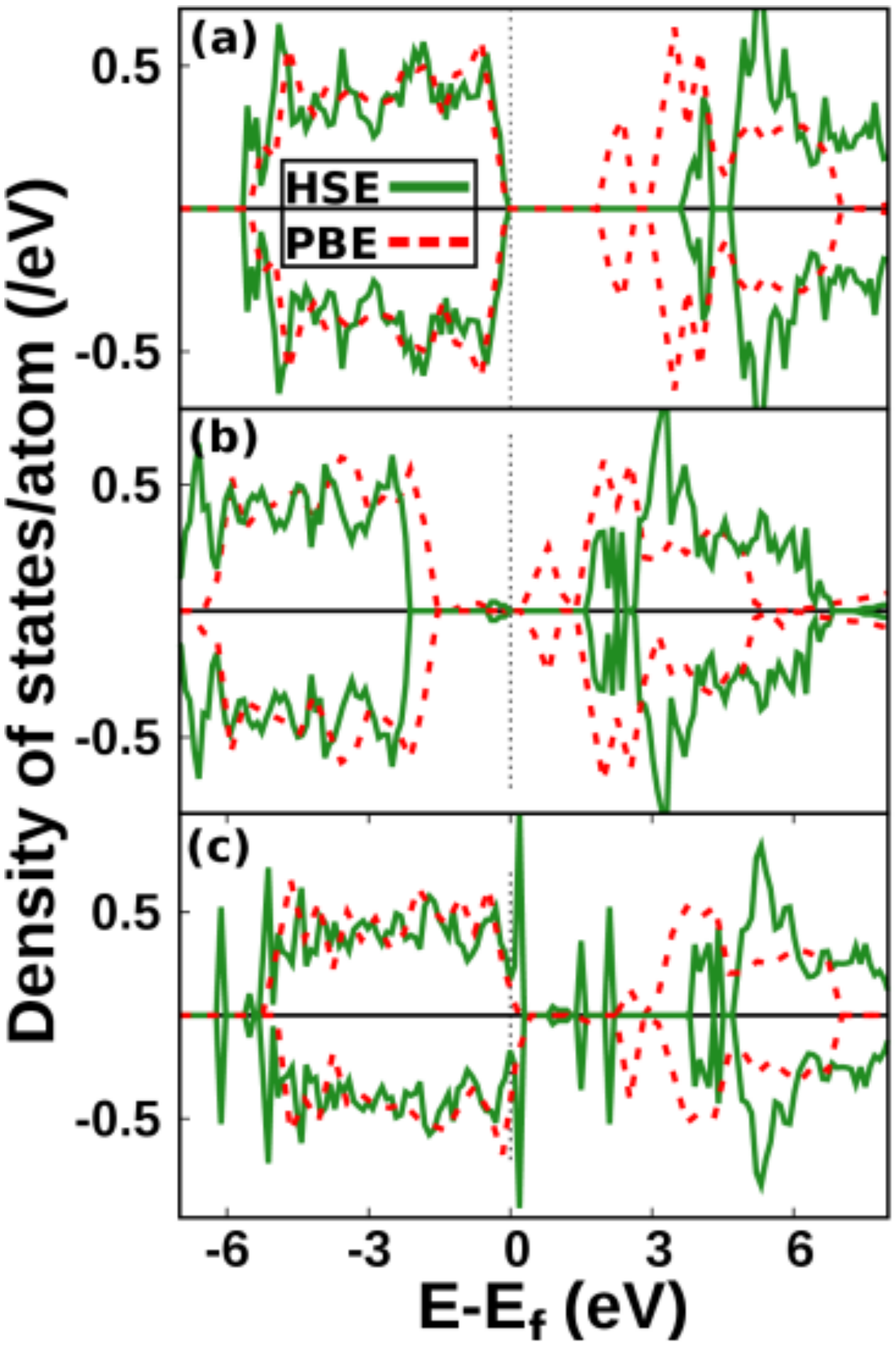

**Figure S8**. (a)-(c) Total density of states calculated with PBE (red dashed line) and HSE (green solid line) functionals for the cases of pure $V_2O_5$, $V_2O_{4.9}$ i.e. with O-vacancy and O-deficient $V_2O_{4.9}$ with 30% $V^{5+}$ ions replaced with $V^{4+}$ ions in the vertical manner, respectively.

## H. Magnetic Susceptibility vs field analysis

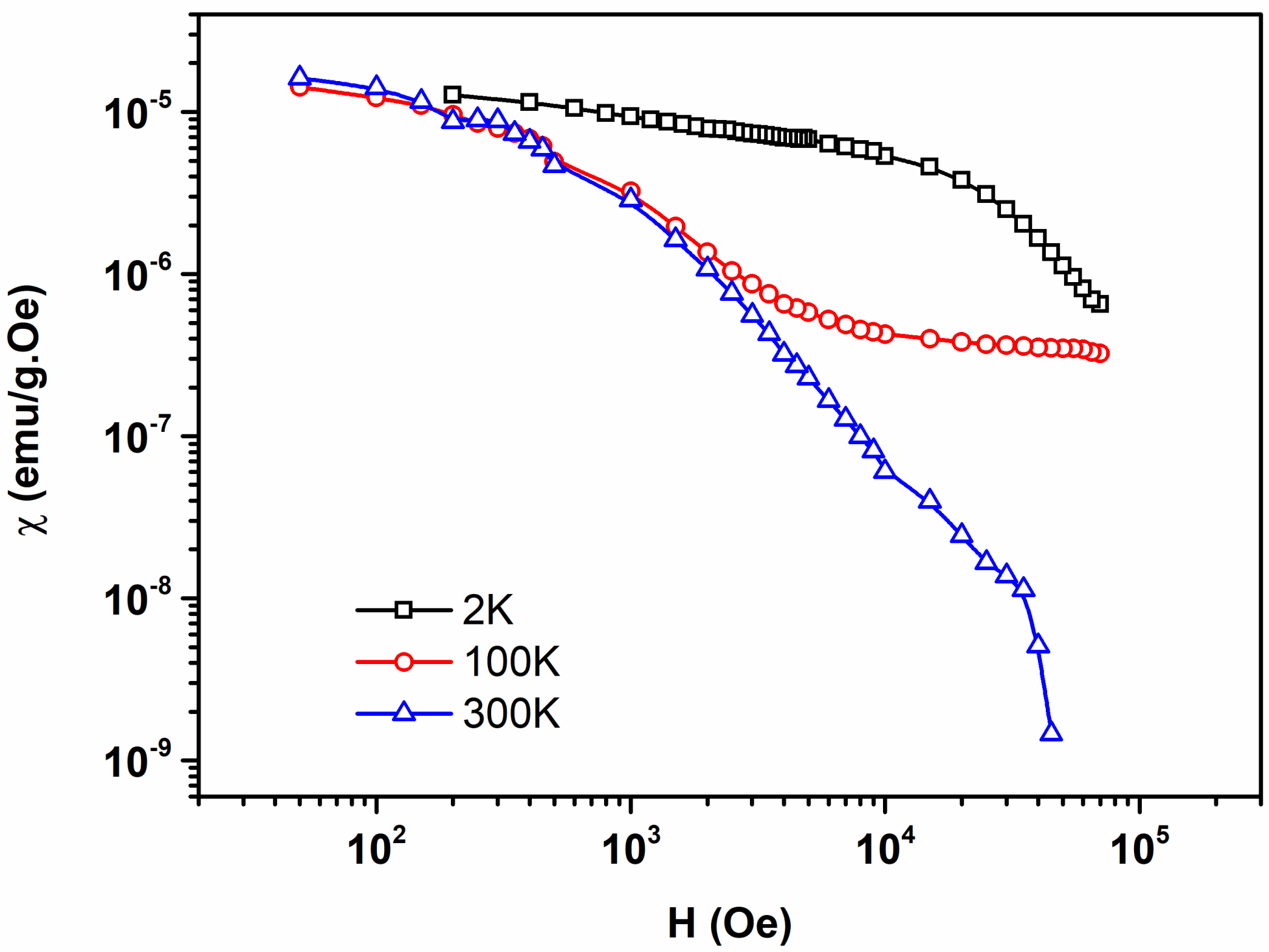

**Figure S9**. The log χ vs log H plot shoring field dependent χ variations at 2, 100 and 300 K.

A flat region log χ vs log H plot depicts a constant susceptibility over decades of field value which implies a linear monotonous rise in χ with filed like a PM or AFM. Whereas a rapidly falling χ value with H represents saturation in moment (M) and hence no further rise in χ with H is possible. Here. The plots show two distinct regions of χ variations with H values indicating different (competing) magnetic interactions at play subjected to field values.